\documentclass[CRMATH,Unicode,manuscript]{cedram}       

\usepackage{hyperref}

\usepackage[utf8]{inputenc} 
\newcommand{\R}{\mathbb{R}}  

\newcommand{\Z}{\mathbb{Z}}
\newcommand{\F}{\mathcal{F}}
\newcommand{\mL}{\mathbb{L}}
\newcommand{\defe}{\stackrel{def.}{=}}
\newtheorem{hypo}{Hypothesis}
\usepackage{todonotes}

\title{Spatial and color hallucinations in a mathematical model of primary visual cortex}

\author{\firstname{Olivier} \middlename{D.} \lastname{Faugeras} \CDRorcid{0000-0001-8363-1550}}
\address{Université Côte d’Azur, Inria,
\\
2004, route des Lucioles - BP 93,
\\
06902 Sophia Antipolis Cedex, France}
\email[O. Faugeras]{olivier.faugeras@inria.fr}
\author{\firstname{Anna}  \lastname{Song}}
\address{Department of Mathematics, Imperial College London, London, UK}
\address{Haematopoietic Stem Cell Laboratory, The Francis Crick Institute, London, UK}
\email[A. Song]{a.song19@imperial.ac.uk}
\author{\firstname{Romain} \lastname{Veltz} \CDRorcid{0000-0003-4653-1475}}
\addressSameAs{1}{}
\email[R. Veltz]{romain.veltz@inria.fr}



\begin{abstract} 
We study a simplified model of the representation of colors in the primate primary cortical visual area V1. The model is described by an initial value problem related to a Hammerstein equation. The solutions to this problem represent the variation of the {activity of populations of neurons in V1} as a function of space and color. The two space variables describe the spatial extent of the cortex while the two color variables describe the hue and the saturation represented at every location in the cortex. We prove the well-posedness of the initial value problem. We focus on its stationary, i.e. independent of time, and periodic in space solutions. We show that the model equation is equivariant with respect to the direct product $\mathcal G$ of the group of the Euclidean transformations of the planar lattice determined by the spatial periodicity and the group of color transformations, isomorphic to $O(2)$, and study the equivariant bifurcations of its stationary solutions when some parameters in the model vary. Their variations may be caused by the consumption of drugs and the bifurcated solutions may represent visual hallucinations in space and color. Some of the bifurcated solutions can be determined by applying the Equivariant Branching Lemma (EBL) by determining the axial subgroups of $\mathcal G$. These define bifurcated solutions which are invariant under the action of the corresponding axial subgroup. We compute analytically these solutions and illustrate them as color images. Using advanced methods of numerical bifurcation analysis we then explore the persistence and stability of these solutions when varying some parameters in the model. We conjecture that we can rely on the EBL to predict the existence of patterns that survive in large parameter domains but not to predict their stability. On our way we discover the existence of spatially localized stable patterns through the phenomenon of "snaking".
\end{abstract}

\begin{document}
	
	\maketitle

	
	\selectlanguage{english}

	\section{Introduction}
		
	Neural Fields are a useful mathematical formalism for representing the dynamics of cortical areas at a macroscopic level, see the reviews \cite{ermentrout:98,bressloff:12,coombes-graben-etal:14}. This formalism has been broadly used to account for the observed activity of the cortical visual area V1. V1 receives inputs from the retina through the LGN. There is a massive feedback from V1 to the LGN. V1 also sends inputs to higher level cortical visual areas such as V2 and V4 and receives feedback signals from them. For a very accessible introduction to the various visual areas, the reader is referred to the book of David Hubel \cite{hubel:95}, and to \cite{kandel-jessell-etal:13} for a more recent presentation. 	
	
	Neural Fields models of V1 are sets of integro-differential equations whose solutions are meant to describe its spatio-temporal activity. The well-posedness of these equations has been studied in depth by various authors \cite{amari:77,faugeras-veltz-etal:09,veltz-faugeras:10,potthast-beim-graben:10}  with a special attention to the stationary solutions, i.e. those which do not depend upon time. These solutions, also called persistent states, are interesting because they appear to provide good models of memory holding tasks on the time scale of the second \cite{colby-duhamel-etal:95,funahashi-bruce-etal:89,miller-erickson-etal:96}. Moreover, they appear to resonate with the fascinating phenomenon of visual hallucinations and their relation with the functional architecture of the visual cortex \cite{ermentrout-cowan:79,bressloff-cowan-etal:01} .
	
	With no exception to our knowledge, all previous work in neural fields theory has not taken into account the chromatic aspects of visual perception. In \cite{song-faugeras-etal:19}, we introduced a neural field model for color perception to explore how the synergy of two antagonistic phenomena, simultaneous contrast and chromatic assimilation, could lead to a "color sensation".

	In the present article, we address the questions of how this model can predict visual hallucinations and what are their spatial and chromatic structures. We are guided in this venture by our previous work \cite{veltz-chossat-etal:15} and make good use of the theory of equivariant bifurcations.
	
	It is structured as follows. In Section \ref{sec:model} we recall the neural field model described in \cite{song-faugeras-etal:19} and prove its well-posedness, Section \ref{sect:stationary} introduces the notion of stationary solutions and their bifurcations.  Section \ref{section:Wspectrum} is dedicated to the computation of the spectrum of the linear operator in the neural field model. Section  \ref{sec:bifurcations} describes the symmetries of the model and uses the equivariant branching lemma to predict the type of bifurcations at the primary bifurcation points and the shape of the bifurcated solutions, or planforms.  Section \ref{sect:explanf} shows examples of such planforms. Section \ref{sec:hallucinations} goes away from this local analysis and explores numerically, thanks to the development of innovative software, a much larger volume of the set of stationary solutions to the neural field equations. We conclude in Section \ref{sect:conclusion}. \	
	\section{The model}\label{sec:model}
	In this work we think of V1 as the closure of a regular domain, noted $\Omega_s$, $s$ stands for space, of $\R^2$, in effect the open square $(-l/2,l/2) \times (-l/2,l/2)$, where $l$ is a positive number which, for simplicity and without loss of generality, we  take equal to 1, except for some of the numerical experiments presented in Section~\ref{sec:hallucinations}. 
	
	The visual cortex is organized into hypercolumns, groups of neurons sharing the same receptive field in the retina and coding for specific physical quantities such as edge orientation, spatial frequency, temporal frequency. These signals are mapped from the retina to V1 following an approximately log polar retinotopic transformation (see Remark~\ref{rem:logpolar}). Unlike in the case of orientation, for which the existence of such hypercolumns in V1 is now well established \cite{tso-zarella-etal:09}, the anatomical and physiological bases for a functional architecture encoding color are still debated. These bases are most likely connected  to the presence of blobs \cite{hubel:86,hubel:95}. Hence, in light of the promising findings made by \cite{xiao-casti-etal:07,chatterjee-ohki-etal:21} it is reasonable to assume in our work a hypercolumnar organisation of cells tuned to a continuum of colors. Our work also supposes the presence of long-range lateral connections between hypercolumns, in agreement with observations of \cite{lund-angelucci-etal:03} where horizontal connections tend to link blobs to blobs. Note that this visual information is stored in the cortex in three dimensions, i.e. the cortex has some thickness. We neglect in our work this thickness and consider only its spatial extent.
	
	We now briefly recall the model described in \cite{song-faugeras-etal:19}. It is based on an opponent representation of colors such as Hering's opponent space \cite{hering:64}. In this setting, a color is a pair $c:=(c_1,c_2)$ of real numbers which encode the \emph{chromaticity} of the color\footnote{We do not consider the achromatic component of a color, except for display, see Section~\ref{subsec:colorequilib}, and restrict ourselves to the chromatic components.}. Details are provided in Appendix \ref{section:color space}. The reader can think of $c_1$ as  encoding the yellow-blue colors and $c_2$  as encoding the red-green colors in Hering's theory. What is important for us is that the set of chromaticities is symmetric w.r.t the origin, i.e. if $c=(c_1,c_2)$ is a chromaticity, then $-c=(-c_1,-c_2)$ is also a chromaticity, called the \emph{opponent} chromaticity or color of $c$. The set of chromaticities is therefore a bounded regular domain of $\R^2$ which is symmetric w.r.t the origin, in effect the open disk $D(0,c_0)$ centered at the origin and of radius $c_0$, where $c_0$ is a positive number which we take without loss of generality and for convenience equal to 1. 
\begin{remark}
We implicitly assume that the topology of the chromaticity space is that of the Euclidean plane. This is only a coarse approximation. Note that the problem of defining a metric in color space is still open \cite{wyszecki-stiles:67,koenderink:10,provenzi:20,berthier:20,vattuone-wachtler-etal:21}.
\end{remark}
	For technical reasons, we consider only the open disk minus a radius, i.e. $D(0,1) \backslash \{[0,1) \times \{0\}\}$. We use polar coordinates to parametrize $D(0,1)$. A chromaticity is therefore represented by a pair $(\rho,\varphi)$ with $\rho \in (0,1)$ and $\varphi \in (0,1)$, keeping in mind that the actual polar coordinates are $(\rho,2\pi \varphi)$.
	\begin{remark}
		The elimination of the semi open radius $[0,1) \times \{0\}$ from $D(0,1)$ is practically not important since the functions that we will manipulate, in particular the eigenfunctions of the operator $W_c$, see below, will be smooth and therefore can be continuously extended to the closed disk $\bar{D}(0,1)$.
	\end{remark}
	The perceptual interpretation of $\rho$ is the saturation of the color $c$ while $\varphi$ is its hue. The opponent chromaticity of $(\rho,\varphi)$ is therefore $(\rho,\varphi+\frac{1}{2} \mod 1)$. The set $\Omega_c$ of chromaticities is therefore the open square $(0,1) \times (0,1)$.

	We define
	\[
	\Omega = \Omega_s \times \Omega_c
	\]
	to be the bounded domain, in effect the open rectangle of $\R^4$, encapsulating the spatial and chromatic coordinates that will be of interest in the sequel.
	
	\subsection{Connectivity kernel}
	Putting all this together, at each point $(r,c)=(r_1,r_2,\rho,\varphi)$ of $\Omega$, we consider a neural mass\footnote{A neural mass is an aggregate of neurons in which the spatial dimension is ignored, see e.g. \cite{wilson-cowan:72}.} whose average membrane potential is noted $V(r,c,t)$.
	It is a function defined on $\Omega \times \mathcal{J}$, $\mathcal{J}$ being an interval of $\R$ containing 0. In \cite{song-faugeras-etal:19}, we assumed that the function $V$ was the solution to an initial value problem\footnote{More precisely we used instead of $V$ the "activity" variable $a=\Upsilon(V)$, see the definition  \eqref{eq:sig} of the sigmoid function below. The two formulations are equivalent in the case $I_{\rm ext}=0$ considered here.} related to a Hammerstein equation\footnote{In mathematical neuroscience this integro-differential equation is often called a Wilson-Cowan equation \cite{wilson-cowan:73} }which writes
	\begin{equation}\label{eq:wcc}
		\tau \frac{\partial V}{\partial t}(r,c,t) = -V(r,c,t)+w \star \Upsilon\left(V (r,c,t)\right)+I_{\rm ext}(r,c,t)   ,
	\end{equation}
	together with the initial condition $V(r,c,0)=V_0(r,c)$.
	
	This equation 
	describes the time variation of the scalar function $V(t)$ defined on $\Omega$ starting from the initial condition $V_0$. At each time $t$, $V(t)$ belongs to some functional space, in effect a Hilbert space $\F$, that we describe in the next section.

	We now discuss the various elements that appear in this equation.
	
	$\tau$ is a time constant that defines the speed of the exponential decay toward the initial condition. Without loss of generality we can  assume $\tau=1$.
	
	$\Upsilon$  is a sigmoidal function mapping $\R$ to the open interval $(0,1)$. It is called the activation function, relating the values of the membrane potential $V$ to the neuronal activity $a$ (0 meaning quiet, 1 meaning highly active).  It writes
	\begin{equation}\label{eq:sig}
		\Upsilon(x)= {\rm Sig}(\gamma x-\varepsilon)\quad \text{  where  }\quad {\rm Sig}(x)=\frac{1}{1+e^{-x}}.
	\end{equation}
	$\varepsilon$ is a parameter that allows us to shift the origin, $\gamma$ controls the slope of the sigmoid at the (shifted) origin, it is often called the nonlinear gain.

	$I_{\rm ext}$ is a function representing the input to the neural mass from different brain areas. In the remaining of this paper we take $I_{\rm ext}=0$, i.e. we consider that area $V1$ is isolated from the rest of the brain. This is clearly an approximation but allows us to do some mathematics and  is a first step toward the analysis of the general case.
	
	$w$ is the connectivity kernel\footnote{It is only qualitatively the same as the one in \cite{song-faugeras-etal:19}.}. It models the influence of neighboring neural masses at $(r',c')$ on the neural mass at $(r,c)$ as a (separable in space and color) linear superposition operation
	\begin{equation}\label{eq:wstar}
		w \star s \left(V (r,\rho,\varphi,t)\right) = \int_{\Omega_s \times \Omega_c} w(r,\rho,\varphi,r',\rho',\varphi')  \Upsilon \left( V(r',\rho',\varphi',t) \right) \,\rho' d\rho'\,d\varphi'\,dr',
	\end{equation}
	with
	\[
	w(r,c,r',c') = w_s(r-r')w_{c}(\rho,\rho',\varphi-\varphi'),
	\]
	where the index $s$ stands for ``space'' and the index $c$ stands for ``chromaticity''.

	If $w(r,c,r',c')$ is positive (respectively negative)  the neural mass at $(r',c')$ excites (respectively inhibits) the one at $(r,c)$. The product of $w_s$ and $w_c$ is intended to model the antagonistic effects of color assimilation and contrast which are parts of the class of perceptual phenomena called chromatic interactions \cite{ware-cowan:82}.
	
	$w_s$ is a ``classical'' two-dimensional ``Mexican hat'' function, see \cite{bressloff:12} and Figure~\ref{fig:jspatial-bif1}-Left, which we write as the difference of two circularly symmetric Gaussians centered at 0: 
	\begin{equation}\label{eq:g}
		w_s(r) = \mu_s e^{-\| r \|_2^2/2\alpha_s^2}-\nu_s e^{-\| r \|_2^2/2\beta_s^2},
	\end{equation}
	where $\|\ \|_2$ is the usual Euclidean $\mathbb{L}_2$ norm in $\R^2$. 
	Biology dictates that $\alpha_s$, $\beta_s$ 
	are very small w.r.t $1$, the size of $\Omega_s$. This indicates that our model only takes into account the  neural connections which are local to the visual area V1 and take place within the gray matter while it is known that different visual areas communicate through the fiber bundles (in the biological sense) forming part of the white matter. This would be part of the term $I_{\rm ext}$ in \eqref{eq:wcc} which we have taken to be equal to 0.

	If $w_s(0) > 0$, i.e. if $\mu_s > \nu_s$, the neural masses at $r'$ such that $\|r - r'\|_2$ is small enough excite the neural mass at $r$, and if $\alpha_s < \beta_s$  those sufficiently far away inhibit it.
	
	We furthermore assume
	\begin{equation}\label{eq:balanced}
		\int_{\Omega_s} w_s(r)\,dr = 0, 
	\end{equation}
	i.e. that the spatial excitation and inhibition are balanced. This is both compatible with some biological evidence (balanced networks \cite{deneve-machens:16}) and mathematically convenient.

	The function $w_c$ is separable in polar coordinates,
	\begin{equation}\label{eq:whpolar}
		w_c(\rho,\rho',\varphi)=w_{c,m}(\rho,\rho') w_{c,a}(\varphi),
	\end{equation}
	with
	\begin{equation}\label{eq:whm}
		w_{c,m}(\rho,\rho') = e^{-\xi_c |\rho^2-\rho'^2|} ,
	\end{equation}
	and
	\begin{equation}\label{eq:wha}
		w_{c,a}(\varphi)=\mu_c e^{-2\pi \alpha_c |(\varphi -\frac{1}{2} \mod 1)-\frac{1}{2}|}- \nu_ce^{-2 \pi \beta_c |(\varphi \mod 1) -\frac{1}{2}|}
	\end{equation}
	The parameters $\xi_c$, $\mu_c$, $\nu_c$,
	$\alpha_c$ and $\beta_c$ are positive and such that $w_{c,a}(c,c')$ is positive if
	$c'$ is close to $c$ and negative if $c'$ is close to $-c$, i.e. in terms of polar coordinates if $\rho'$ is close to $\rho$ and either $\varphi'$ is close to $\varphi$ or $\varphi'$  is close to $\varphi+\frac{1}{2} \mod 1$.
\begin{figure}[h]\label{fig:wca}
\includegraphics[width=0.75\textwidth]{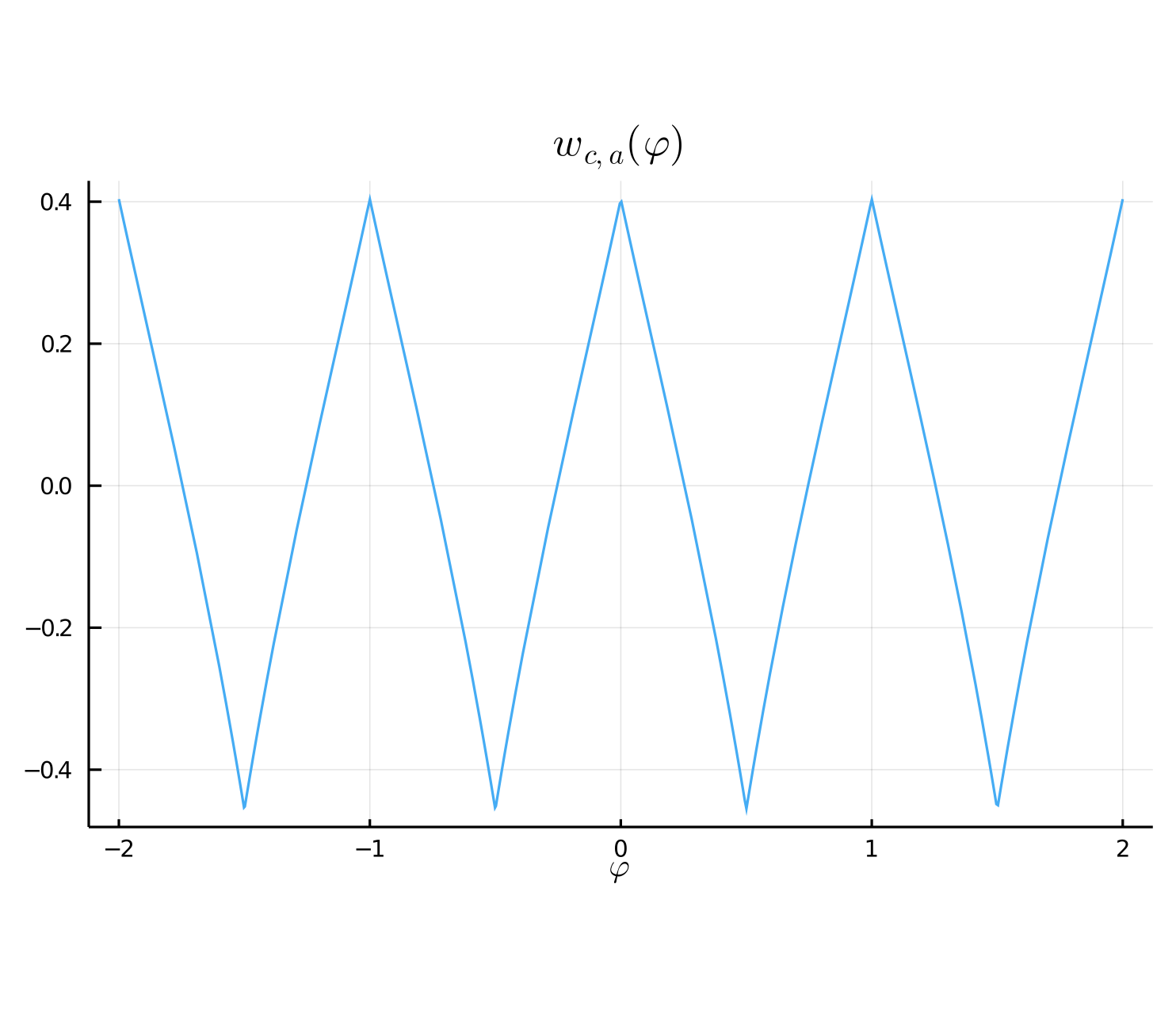}
\caption{Plot of $w_{c,a}(\varphi)$ defined in \eqref{eq:wha} for the values of the parameters 
$\alpha_c = 0.3$, $\beta_c =0.4$, 
		$\mu_c  = 0.6$ and $\nu_c = 0.69$.
}
\end{figure}

	\begin{remark}
	Note that the function $w_{c,a} : \R \to \R$ is 1-periodic and even.
	\end{remark}
	\begin{remark}
		For technical convenience we perform the change of variable $\rho^2 \to \varrho$ so that $w_{c,m}$ simplifies to
		\begin{equation}\label{eq:whmsquare}
			w_{c,m}(\varrho,\varrho')=e^{-\xi_c|\varrho-\varrho'|},
		\end{equation}
		and the measure $\rho\,d\rho$ on $(0,1)$ is equal to $\frac{1}{2}d\varrho$. A chromaticity is therefore represented by a pair $(\varrho,\varphi)$, $\varrho,\,\varphi \in (0,1)$.
	\end{remark}

	The product $w$ of $w_s$ and $w_c$ behaves as a function of $r,\,r',\,c,\,c'$ as shown in Table \ref{table:sign} adapted from \cite{song-faugeras-etal:19}.
	\begin{table}[ht]
		\caption{\textbf{Sign of the connectivity kernel $w$}}
		\begin{tabular}{| c | c | c |}
			\hline
			$w_s(r-r')w_c(\varrho,\varrho',\varphi-\varphi')$ & $c'$ close to $c$ & $c'$ close to $-c$ \\
			\hline
			$r'$ close to $r$ & $> 0$ & $< 0$\\
			\hline
			$r'$ far from $r$ & $< 0$ & $> 0$ \\
			\hline
		\end{tabular}
		\label{table:sign}
	\end{table}
	As pointed out in this paper, this is in qualitative agreement with the nonlinear behaviour of color shifts found in \cite{monnier-shevell:04,monnier:08}.
	
	Note that by definition of $w_s$ and $w_c$ the function $w$ is symmetric w.r.t. $r$ and $r'$, and $c$ and $c'$, respectively.

	\subsection{Choice of the appropriate functional space $\F$ and well-posedness of \eqref{eq:wcc}}
	Our choice of $\F$ is guided by three criteria: 
	\begin{enumerate}
		\item The well-posedness of the problem,
		\item Its biological relevance, 
		\item Its suitability for numerical computations.
	\end{enumerate}
	
	The choice of a Hilbert space is appealing and a natural choice is $\F = \mL^2(\Omega)$. As argued in \cite{veltz-faugeras:10}, this unfortunately allows the membrane potential to be singular since for example the function $(r,c) \to \|r-(\frac 12,\frac 12)\|_2^{-1/2}\ \in \mL^2(\Omega)$. It is desirable that the average membrane potential stays bounded on the cortex and a way to achieve this is to allow for more spatial and chromatic regularity by assuming that $(r,c) \to V(r,c)$ is differentiable almost everywhere.
	
	The choice of the Sobolev space $\F=W^{m,2}$, is convenient for two reasons. First it is a Hilbert space endowed with the usual inner product:
	\begin{equation}\label{eq:inner}
		\langle V_1,V_2 \rangle_\F = \sum_{|\alpha|=0}^m \langle D^\alpha V_1,\,D^\alpha V_2 \rangle_{\mL^2(\Omega)}\quad \forall V_1,\,V_2 \in \F
	\end{equation}
	where the multi-index $\alpha$ is a sequence $\alpha=(\alpha_1,\alpha_2,\alpha_3,\alpha_4)$ of 4 integers and  $|\alpha|=\sum_{i=1}^4 \alpha_i$, and the symbol $D^\alpha$ represents a partial derivative, e.g.,
	\[
	D^\alpha V =\frac{\partial^{\alpha_1+\cdots+\alpha_4}}{\partial r_1^{\alpha_1}\partial r_2^{\alpha_2}\partial \varrho^{\alpha_3}\partial \varphi^{\alpha_4}}V
	\]
	The second reason is that, because the boundary of $\Omega$  is sufficiently regular (it satisfies the cone-property \cite{adams:75, adams-fournier:03}), $\F$ is a commutative Banach algebra for pointwise multiplication \cite{adams:75}[Chapter V, Theorem 5.23]. 
	
This is necessary in the upcoming bifurcation analysis in order to apply the Equivariant Branching Lemma, see Proposition~\ref{prop:ebl}, for which we require some smoothness of $w$.	\begin{remark}
		The reader probably wonders what is the value of $m$. In order to have the Banach algebra property, we need to have $2m > d$, $d$ being the dimension of $\Omega$, i.e. 4. Hence the smallest possible value of $m$ is 3. But in Section \ref{sec:hallucinations} we  assume that $\Omega_c$ is one-dimensional making $d$ equal to 3 and the smallest possible value of $m$ is equal to 2 in this case.
	\end{remark}
	
	We next define the operator $W$ as acting on $\mL^2(\Omega)$ as follows. Let 
	$U \in L^2(\Omega)$, we define
	\[
	W \cdot U(r,c) :=  w \star U(r,c)= \int_{\Omega}  w_s(r-r')w_c(c,c') U(r',c') \,dr' \,dc'
	\]
	\begin{remark}
		Remember that the measure $dc'$ is in effect equal to $\frac{1}{2}d\varrho'\,d\varphi'$.
	\end{remark}
	It is clear that this is well defined and 
	we have the following proposition.
	\begin{proposition}
		The operator $W$ maps $\mL^2(\Omega)$ to $\F$ and hence $\F$ to $\F$.
	\end{proposition}
	\begin{proof}
		The proof follows from Proposition 2.3 in \cite{veltz-faugeras:10}.
	\end{proof}
	
	We also have the following Proposition about the solutions to \eqref{eq:wcc}.
	\begin{proposition}
		For each $V_0 \in \F$ there exists a unique solution in $\mathcal{C}(\R^+, \F)$ to the following Cauchy problem
		\begin{equation}\label{eq:wcc1}
			\left\{ \begin{array}{ccl}
				\frac{dV}{dt} &=& -V + W \cdot \Upsilon(V)\\
				V(0) & = & V_0,
			\end{array}
			\right.
		\end{equation}
		where the function $s$ is defined by \eqref{eq:sig}.
	\end{proposition}
	\begin{proof}
		The proof  is a an immediate consequence of Proposition 2.5 in \cite{veltz-faugeras:10}.
	\end{proof}
	
	\section{Stationary solutions and bifurcations thereof}\label{sect:stationary}
	
	In this paper, we focus on the stationary (independent of time) solutions to \eqref{eq:wcc1}, the steady-states. They are important because they are thought to be good models of the memory holding tasks on the timescale of the second as demonstrated experimentally on primates \cite{colby-duhamel-etal:95,funahashi-bruce-etal:89,miller-erickson-etal:96}. These solutions may change drastically when some of the parameters such as $\gamma$ and $\varepsilon$ in \eqref{eq:sig}, $\mu_s$, $\nu_s$, $\alpha_s$, $\beta_s$ in \eqref{eq:g}, $\xi_c$ in \eqref{eq:whm}, and $\mu_c$, $\nu_c$, $\alpha_c$, $\beta_c$ in \eqref{eq:wha} vary. In effect, we will concentrate on the first  parameter, the nonlinear gain $\gamma$, which is important in determining the relation between neuronal activity (a number between 0 and 1) and the associated membrane potential $V$. It is known that the ingestion of drugs such as LSD and marijuana a) can change this relation and b) can trigger hallucinatory patterns \cite{oster:70,siegel:77}. It is therefore very much worth our efforts to investigate if, when varying the parameter $\gamma$, stationary solutions to \eqref{eq:wcc1} do bifurcate and if  the bifurcated solutions resemble some of the known visual illusions.

	To summarize, we are going to study the bifurcations when $\gamma$  varies of the solutions  to the  equation
	\begin{equation}\label{eq:wccs}
		F(V,\gamma)=0,
	\end{equation}
	where the operator $F$ is defined by
	\begin{equation}\label{eq:F}
		F\defe\left\lbrace
		\begin{array}{ccc}
			\mathbb{W}^{m,2}(\Omega)\times \R & \longrightarrow& \mathbb{L}^2(\Omega)\\
			(V,\,\gamma)&\longrightarrow&  -V+W \cdot \Upsilon\left(V \right).
		\end{array}\right.
	\end{equation}
	In order to achieve such a task, we need to determine the spectrum of the operator $W$ and the symmetry properties of the operator $F$  with respect to some groups of transformations of $\Omega$.
	We describe the spectrum of $W$ in Section~\ref{section:Wspectrum} and the symmetry properties of $F$ in Section~\ref{sec:bifurcations}.

	It is natural at this point to introduce the two operators $W_s$ and $W_c$ acting on
	$\mL^2(\Omega)$ 
	associated to the functions $w_s$ of \eqref{eq:g} and $w_c$ of \eqref{eq:whpolar} as follows
	\[
	W_s \cdot U(r,c) = \int_{\Omega_s} w_s(r-r') U(r',c)\,dr'\quad \text{and} \quad W_c \cdot U(r,\varrho,\varphi) = \frac{1}{2}\int_{\Omega_c} w_c(\varrho,\varphi,\varrho',\varphi') U(r,\varrho',\varphi')\,d\varrho'\,d\varphi'.
	\]
	The separability of the function $w$ in the space and color variables is reflected in the notation $W = W_c \otimes W_s$. Fubini's Theorem dictates that $W$ is also equal to $W_s \otimes W_c$.
	
	Similarly the separability of the function $w_c$ as the product of $w_{c,m}$ in \eqref{eq:whm} for the magnitude and $w_{c,a}$ in \eqref{eq:wha} for the angle implies that $W_c$ is itself separable, $W_c=W_{c,m} \otimes W_{c,a} =  W_{c,a} \otimes W_{c,m} $:
	\begin{equation}\label{eq:whmwha}
		W_{c,m} \cdot U(r,\varrho,\varphi) = \frac{1}{2}  \int_0^1 w_{c,m}(\varrho,\varrho')U(r,\varrho',\varphi)\,d\varrho' \quad \text{and} \quad W_{c,a} \cdot U(r,\varrho,\varphi)= \int_0^1 w_{c,a}(\varphi-\varphi') U(r,\varrho,\varphi')\,d\varphi'
	\end{equation}
	where the convolution in the right hand side of the definition of $W_{c,a}$ is a periodic convolution.

	\section{The spectrum of $W$ in $\mathbb{L}^2(\Omega)$}\label{section:Wspectrum} 
	The reader can verify that, given the symmetry properties of the functions $w_c$ and $w_s$, the operator $W$ is symmetric in $\mathbb{L}^2(\Omega)$, i.e. satisfies
	\[
	\langle U_1,\,W \cdot U_2 \rangle_{\mathbb L^2(\Omega)} = \langle W  \cdot U_1,\, U_2 \rangle_{\mathbb L^2(\Omega)}.
	\]
	
	Another important property of $W$ is that it is {\emph compact}, i.e. given any bounded sequence $(U_n)_{n \geq 0}$ of functions in $\mathbb{L}^2(\Omega)$ the sequence $(W \cdot U_n)_{n \geq 0}$ contains a converging subsequence for the norm of $\mathcal F$. This is a direct consequence of  Lemma 2.4 in \cite{veltz-faugeras:10}.
	
	A consequence of the fact that $W$ is compact is that its spectrum is compact and at most countable. Moreover each point in the spectrum, except perhaps 0, is isolated. All the non zero elements of the spectrum are eigenvalues and all eigenvalues are real since $W$ is symmetric.
	
	Since $w$ is separable in space and color, the spectrum of $W$ is defined by the spectra of $W_s$ and $W_c$: the eigenvalues are the product of those  of each operator and the eigenfunctions are separable in space and color, being the products of those of $W_s$ and $W_c$.
	\subsection{The spectrum of $W_s$ in $\mathbb{L}^2(\Omega_s)$}\label{subsect:wsspectrum}
	Because we are studying the symmetry-breaking steady-state bifurcations of the solutions to our model,  we know from previous work on heat-conduction in fluids \cite{chossat-lauterbach:00} or on visual hallucinations \cite{bressloff-cowan-etal:01} that this leads to the formation of spatially periodic patterns.
	
	This brings into play the lattice $\mathcal{L}$ generated by the two vectors $k_1$ and $k_2$ where $(k_1,k_2)$ is the canonical basis of $\R^2$ and commands that we quotient $\R^2$ by $\mathcal{L}$ thus obtaining the 2-torus $\tilde{\Omega}_s=\R^2/\mathcal{L}$. The dual lattice $\mathcal{L}^*$ is also generated by the two vectors $k_1$ and $k_2$.

	We thus make the following assumption about the solutions to equations \eqref{eq:wcc} which is inspired by biology with the advantage that it simplifies the mathematics:
	\begin{hypo}
		The biologically relevant solutions to  \eqref{eq:wcc} are $\mathcal{L}$-periodic i.e. satisfy
		\[
		V(r+\boldsymbol{l},c,t)=V(r,c,t)\quad  \forall \boldsymbol{l} \in \mathcal{L}
		\]
	\end{hypo}
	We now work on the Hilbert space $\mathbb{L}^2(\tilde{\Omega}_s)$ of $\mathcal{L}$-periodic functions with the same inner product as before. We note $\tilde{W}_s$ the operator $W_s$ restricted to this space i.e. defined by
	\[
	\tilde{W}_s \cdot V (r)= \int_{\Omega_s} w_s\left(r-r'\mod \mathcal{L}\right) V(r') dr' \quad \forall r \in \Omega_s,
	\]
	Note that the spatial convolution in \eqref{eq:wstar} is now a \emph{periodic} convolution.
	$\tilde{W}_s$ is clearly a symmetric compact operator on $\mathbb{L}^2(\tilde{\Omega}_s)$.
	
	It is easy to characterize the spectrum of $\tilde W_s$:
	\begin{proposition}\label{prop:specWs}
		The eigenvalues $\lambda^s_{m,n}$, $m,\,n \in \Z$, of $\tilde{W}_s\in \mathbb{L}^2(\tilde{\Omega}_s)$ are the (real) Fourier coefficients of the $\mathcal{L}$-periodic even function $\tilde{w}_s$:
		\[
		\lambda^s_{m,n} = \int_{\Omega_s} w_s (r)\cos  2\pi \langle k_{m,n},\, r \rangle \,dr,
		\]
		where
		\[
		k_{m,n}\defe mk_1+n k_2.
		\]
		Because $w_s$ is even, so is the sequence of eigenvalues: $\lambda^s_{m,n}=\lambda^s_{-m,-n}$, $m,\,n \in \Z$.
		For $(m,n) \neq (0,0)$ the eigenspaces of $\tilde{W}_s$ are of even dimension, larger than or equal to 2. 
		If $(m,n)=(0,0)$, because of \eqref{eq:balanced}, $\lambda^s_{0,0}=0$ and the dimension of the kernel of $\tilde{W}_s$ is 1.
		 Given an eigenvalue $\lambda^s_{m,n}$ the corresponding eigenspace is generated by the functions:
		\[
		\cos 2 \pi \langle k_{p,q},\, r \rangle \quad  \text{and} \quad \sin 2 \pi\langle k_{p,q},\, r \rangle,
		\]
		for all $(p,q)$ such that $\lambda^s_{p,q}=\lambda^s_{m,n}$.
	\end{proposition}
	\begin{proof}
		This follows for example from the fact that $\tilde{W}_s$ is diagonalized by the Fourier Transform.
	\end{proof}

	\subsection{Spectrum of $W_c$ in $\mathbb{L}^2(\Omega_c)$}\label{subsect:whspectrum}
	Define the operator $\tau_c$ acting on the $\varphi$-periodic functions of period $1$ as
	\begin{equation}\label{eq:Rpdef}
		\tau_c \cdot u(\varrho,\varphi) = u(\varrho , -\varphi \mod 1)
	\end{equation}
	We have the following Proposition.
	\begin{proposition}\label{prop:opS}
		The linear operator $W_c$ on $\mathbb{L}^2(\Omega_c)$ is symmetric and compact, it commutes with the operator $\tau_c$ defined by \eqref{eq:Rpdef}. 
	\end{proposition}
	\begin{proof}
		Obvious from the definitions.
	\end{proof}
	From this we have
	\begin{corollary}
		The spectrum of $W_c$ is real and at most countable. The eigenfunctions are separable w.r.t. $\varrho$, $\varphi$.
	\end{corollary}
	\begin{proof}
		The first assertion follows from the symmetry and compactness of $W_c$. The second follows from the definitions \eqref{eq:whpolar}-\eqref{eq:wha}.
	\end{proof}
	It remains to characterize the eigenfunctions and eigenvalues.
	\begin{proposition}\label{prop:eigenwha}
		The eigenspaces of $W_{c,a}$ are generated by the functions $\sin 2\pi n\varphi$ and $\cos 2\pi  n \varphi$, $n \in \mathbb{N}$. Hence they are one-dimensional for $n=0$, two-dimensional otherwise. The corresponding eigenvalues are the (real) Fourier coefficients of the $1$-periodic even function  given by \eqref{eq:wha} of index $n$.
	\end{proposition}
	\begin{proof}
		This follows immediatly from the fact that the operator $W_{c,a}$ is a $1$-periodic convolution and the $1$-periodic function $\mu_c e^{-2\pi \alpha_c |(\varphi -\frac{1}{2} \mod 1)-\frac{1}{2}|}- \nu_ce^{-2 \pi \beta_c |\varphi \mod 1) -\frac{1}{2}|}$ is even.
	\end{proof}
	Regarding $W_{c,m}$, we have the following Proposition.
	\begin{proposition}\label{prop:eigenwhm}
		The eigenspaces of $W_{c,m}$ corresponding to the non zero eigenvalues are one-dimensional. Its kernel is reduced to the null function. The  eigenvalues $\lambda_n^c$, $n \in \mathbb{N}$, are obtained from the countable solutions $x_n$ to the transcendental equation
		\begin{equation}\label{eq:xwhm}
			\tan x = \frac{2\xi_c x}{x^2-\xi_c^2}\quad x >0
		\end{equation}
		from the relation
		\begin{equation}\label{eq:eigenvalwhm}
			\lambda_n^c = \frac{\xi_c}{\xi_c^2+x_n^2}
		\end{equation}
		The eigenfunction $e_{\lambda_n}^{c,m}$ corresponding to the eigenvalue $\lambda_n \neq 0$ is given by
		\begin{equation}\label{eq:eigenfuncwmh}
			e_{\lambda_n}^{c,m}(\varrho) = \xi_c \sin  x_n \varrho +  x_n \cos  x_n \varrho
		\end{equation}
	\end{proposition}
	\begin{proof}
		We have, according to \eqref{eq:whmsquare} and \eqref{eq:whmwha}
		\[
		W_{c,m} \cdot u(\varrho)  =\frac{1}{2} \int_0^1  e^{-\xi_c |\varrho-\varrho'|}u(\varrho')\,d\varrho'
		\]
		Deriving twice w.r.t. $\varrho$
		\[
		\left(W_{c,m} \cdot u\right)^{(2)}(\varrho) =\xi_c^2W_{c,m} \cdot u (\varrho)-\xi_c u(\varrho),
		\]
		so that if $u$ is an eigenfunction corresponding to the eigenvalue $\lambda^c$ we obtain 
		\[
		\lambda^c u^{(2)}(\varrho)=\xi_c \left(\xi_c \lambda^c-1\right) u(\varrho),
		\]
		which  shows that  if $\lambda^c = 0$, $u=0$. If $\lambda^c\neq 0$ we define $K:=\xi_c\left(\xi_c-\frac{1}{\lambda^c}\right)$. The eigenfunctions are solutions to
		\begin{equation}\label{eq:oderho}
			u^{(2)}(\varrho)=Ku(\varrho)
		\end{equation}
		which are of the form  $u(\varrho);=a e^{ \sqrt{K} \varrho}+b e^{ -\sqrt{K} \varrho}$ for some constants $a$ and $b$, real or complex. $\lambda^c$, $a$ and $b$ are determined by writing that the function $u$ thus defined is indeed an eigenfunction of $W_{c,m}$ corresponding to the eigenvalue $\lambda^c$.
		
		It can be verified that if $K \geq 0$, i.e. if $\lambda^c \geq \frac{1}{\xi_c}$, there are no solutions. Hence we must assume $\lambda^c < \frac{1}{\xi_c}$ and hence $K < 0$. The solutions to \eqref{eq:oderho} write
		\begin{equation}\label{eq:eigenfWh}
			u(\varrho)=a \sin \sqrt{-K}\varrho + b \cos  \sqrt{-K}\varrho
		\end{equation}
		A symbolic computation system shows that
		\begin{multline*}
			W_{c,m}  \cdot u(\varrho)= \lambda^c u(\varrho) +\\
			\frac{\lambda^c}{2\xi_c}\Big(e^{\xi_c \varrho}e^{-\xi_c}\left(-a\left(\sqrt{-K} \cos \sqrt{-K}+\xi_c \sin \sqrt{-K}\right)+b\left(\sqrt{-K} \sin \sqrt{-K}-\xi_c \cos \sqrt{-K} \right) \right)+\\
			e^{-\xi_c \varrho}\left(a \sqrt{-K}-b\xi_c \right)\Big)
		\end{multline*}
		The two conditions
		\[
		-a\left(\sqrt{-K} \cos \sqrt{-K}+\xi_c \sin \sqrt{-K}\right)+b\left(\sqrt{-K} \sin \sqrt{-K}-\xi_c \cos \sqrt{-K} \right) =0,
		\]
		and
		\begin{equation}\label{eq:ab}
			a \sqrt{-K}-b\xi_c=0
		\end{equation}
		are necessary and sufficient to guarantee that  $ W_{c,m}  \cdot u(\varrho)= \lambda^c u(\varrho)$ for all $0 < \varrho < 1$. Since $a$ and $b$ are not both equal to 0 (otherwise $u=0$ and hence $\lambda^c=0$) we must have
		\[
		2\xi_c \sqrt{-K}  \cos \sqrt{-K}+(\xi_c^2+K) \sin \sqrt{-K} = 0,
		\]
		which is equivalent to
		\[
		\tan \sqrt{-K} =-\frac{2\xi_c \sqrt{-K}}{\xi_c^2+K}
		\]
		Letting $x=\sqrt{-K}$, the eigenvalues are found by solving for $x >  0$ the equation
		\[
		\tan x = \frac{2\xi_c x}{x^2-\xi_c^2}.
		\]
		For each value of $x$, the corresponding value of $\lambda^c$ is found to be
		\[
		\lambda^c = \frac{\xi_c}{\xi_c^2+x^2}
		\]
		The function $x \to  \frac{2\xi_c x}{x^2-\xi_c^2 }$ decreases monotonically from $+\infty$ for $x=\xi_c^+$ to 0 when $x \to +\infty$. The corresponding curve therfore intersects the curve representing the $\pi$-periodic function $x \to \tan x$ at a countable number of points $x_n$, $n \geq 0$, yielding the eigenvalues $\lambda^c_n$ of $W_{c,m}$.
		
		According to \eqref{eq:eigenfWh} and \eqref{eq:ab} the corresponding eigenfunction is
		\[
		e_{\lambda^c_n}^{c,m}(\varrho) = \xi_c \sin  x_n \varrho +  x_n \cos  x_n \varrho
		\]
	\end{proof}
	\section{Symmetries of the model and equivariant bifurcations of the solutions}\label{sec:bifurcations}
	It follows from \eqref{eq:balanced} that $\overline{V}=0$ is a solution to \eqref{eq:wccs} and \eqref{eq:F} for all values of $\gamma$. We are interested in the problem of determining how this solution bifurcates when $\gamma$ increases, while allowing us to change somewhat the value of the threshold $\varepsilon$ in \eqref{eq:sig}. 
	
Because of \eqref{eq:wccs}, \eqref{eq:F} and \eqref{eq:sig} we have
		\[
		dF(0,\gamma)=-{\rm Id}+\gamma {\rm Sig}(-\varepsilon) W,
		\]
		Let $\lambda_p$ be the largest positive eigenvalue of $W$. When one increases $\gamma$ from 0 to $\gamma_p \defe 1/(\lambda_p {\rm Sig}(-\varepsilon))$ the solution $\overline{V}=0$ bifurcates to another solution. 
	
	We  investigate the properties of the operator $F$ defined by \eqref{eq:F} under the action of the Euclidean group $E(2)$ restricted to the lattice $\mathcal{L}$ defined at the beginning of Section~\ref{subsect:wsspectrum} for the spatial part, and the group $O(2)$ for the color part. This group arises from the rotations $R_{\varphi_c}^c$ acting on the $\varphi$ angle by $\varphi \to \varphi+\varphi_c \mod 1$ and the reflection $\tau_c: \varphi \to -\varphi \mod 1$. An element $\psi$ of this group is of the form 
\begin{equation}\label{eq:O2element}
	\psi = R_{\varphi_c}^c \tau_c^m, \, m \in \{0,\,1\}.
\end{equation}

	The action of $E(2)$ on the space of $\mathcal{L}$-periodic functions is best understood by considering separately the translations and the rotations. Since the translations of $\R^2$ leave the set of $\mathcal{L}$-periodic functions invariant and translations in $\mathcal{L}$ fix all $\mathcal{L}$-periodic functions, the effective action of the group of translations of $\R^2$ is as the 2-torus $T^2=\R^2/\mathcal{L}$ which is compact.
	For the rotations, recall that the holohedry $H$ of the lattice $\mathcal{L}$ is the largest subgroup of $O(2)$ that leaves $\mathcal{L}$ invariant. In the case of a square lattice $H=D_4$ is the dihedral group of the symmetries of the square. It follows that the largest subgroup of $E(2)$ that leaves $\mathcal{L}$ invariant is the compact semi-direct sum $E(\mathcal{L}):=D_4 \overset{\cdot}{+} T^2$. Therefore, we are interested in the action on $F$ of the compact group $\mathcal{G}=E(\mathcal{L}) \times O(2)$. We have the following Proposition.
\begin{proposition}\label{prop:equivariant}
The operator $F$ defined in \eqref{eq:F} is equivariant w.r.t. the action of the group $\mathcal{G}$.
\end{proposition}
\begin{proof}
	The action of the element $g=(\varpi,\psi) \in \mathcal{G}$ on $F$, with $\psi$ given by \eqref{eq:O2element}, is defined as follows
	\[
	g \cdot F( V,\gamma) (r,\varrho,\varphi) = F(V,\gamma)(\varpi^{-1}r,\varrho,(-1)^m(\varphi-\varphi_c)),\ m \in \{0,\,1\},
	\]
and, because of Proposition~\ref{prop:opS}, the reader will  verify that $F$ is equivariant w.r.t. the action of $\mathcal{G}$, i.e. that we have
	\begin{equation}\label{eq:invF}
		g \cdot F( V,\gamma) (r,c) = F(g \cdot V,\gamma)(r,c)\ \forall g \in \mathcal{G}.
	\end{equation}
\end{proof}
	At this point, branches of planforms are usually obtained by applying the equivariant branching lemma \cite{vanderbauwhede:80,cicogna:81,golubitsky-stewart-etal:88,chossat-lauterbach:00} as follows. The kernel $\mathcal{V}$ of $dF(0,\gamma_p)$ is $\mathcal{G}$-invariant. We fix an isotropy subgroup $\Sigma$ of $\mathcal{G}$ (i.e. for which there exists  $v_0 \in \mathcal{V}$ such that $g \cdot v_0 = v_0$ for all $g \in \Sigma$) and compute the dimension  of the fixed point subspace ${\rm Fix}_{\mathcal{V}}(\Sigma)$ of $\mathcal{V}$ associated with $\Sigma$ where
	\[
	{\rm Fix}_{\mathcal{V}}(\Sigma)\defe \left \{ v \in \mathcal{V},\, \sigma \cdot v = v, \  \forall \sigma \in \Sigma \right \}.
	\]
	The equivariant branching lemma states that if $\Sigma$ is an axial subgroup of $G$, i.e. such that
	\begin{equation}\label{eq:fix1}
		{ \rm dim }\ {\rm Fix}_{\mathcal{V}}(\Sigma)=1,
	\end{equation}
	then generically, there is a branch of steady-state solutions to \eqref{eq:wcc} with symmetries $\Sigma$. The genericity condition requires that the eigenvalue that goes through 0 does so with nonzero speed and that some components of  the Taylor expansion of $F$ are non zero. 
	
We characterize the isotropy subgroups of $\mathcal{G}$ and the corresponding fixed point subspaces  in the following Proposition.
\begin{proposition}\label{prop:isoG}
The isotropy subgroups $\Sigma$ of $\mathcal{G}=E(\mathcal{L}) \times O(2)$ fall in two classes: those which do not contain the color reflection $\tau_c$ are those of $\mathcal{H} := E(\mathcal{L}) \times SO(2)$, and those that do contain $\tau_c$.  In the first case the corresponding fixed subspaces are those of $\Sigma$ as a subgroup of $\mathcal{H}$. In the second case the fixed subspaces are those of $\Sigma \cap \mathcal{H}$ projected on $\mathcal{V}_{1}$ along $\mathcal{V}_{-1}$, the eigenspaces of $\tau_c$ corresponding to the eigenvalues 1 and -1, respectively.
\end{proposition}
\begin{proof}
If we restrict the color group $O(2)$ to $SO(2)$, i.e. if we only consider the action of the color rotations, our symmetry group is the same as for the equivariant Hopf bifurcation with $E(\mathcal{L})$ symmetry \cite{silber-knobloch:91,dionne-golubitsky-etal:95}. We note that the fixed point subspaces have even dimension since $SO(2)$ commutes with $E(\mathcal{L})$ and all real finite dimensional non  trivial representation  of $SO(2)$ are of even dimension. 
All fixed point subspaces of $\mathcal{V}$ for the action of $\mathcal{H}$ are therefore of even dimension. 

If we now extend $SO(2)$ to $O(2)$, i.e. $\mathcal{H}$ to $\mathcal{G}$, given an isotropy subroup $\Sigma$ of $\mathcal{G}$, either it is an isotropy subgroup  of $\mathcal{H}$ and we are done, or it is not and the color reflection operator $\tau_c$ is not reduced to the identity on ${\rm Fix}_{\mathcal{V}}(\Sigma)$. But $\tau_c$ is a projector. It has therefore two eigenspaces $\mathcal{V}_{\pm 1}$ corresponding to its eigenvalues $\pm 1$ and $\mathcal{V}$ is the direct sum of the corresponding eigenspaces. Let $v \in {\rm Fix}_{\mathcal{V}}(\Sigma \cap \mathcal{H})$. We write $v=v_1+v_{-1}$ where $v_{\pm 1} \in \mathcal{V}_{\pm 1}$. We have
\begin{equation}\label{eq:vpm1}
\tau_c \cdot v = \tau_c \cdot v_1 + \tau_c \cdot v_{-1} = v_1 -v_{-1},
\end{equation}
so that the projection $p_1\left({\rm Fix}_{\mathcal{V}}(\Sigma \cap \mathcal{H})\right)$ of ${\rm Fix}_{\mathcal{V}}(\Sigma \cap \mathcal{H})$ on $\mathcal{V}_1$ along $\mathcal{V}_{-1}$ is included in ${\rm Fix}_{\mathcal{V}}(\Sigma)$. 

Conversely, let $v \in {\rm Fix}_{\mathcal{V}}(\Sigma)$. The condition $v \in {\rm Fix}_{\mathcal{V}}(\Sigma)$ and \eqref{eq:vpm1} impose $\tau_c \cdot v = v$, i.e. $\tau_c \cdot v_{-1} = 0$, i.e. $v_{-1}=0$. so that $v=v_1$ and ${\rm Fix}_{\mathcal{V}}(\Sigma ) \subset p_1\left({\rm Fix}_{\mathcal{V}}(\Sigma \cap \mathcal{H}) \right)$ . 
\end{proof}

\
\begin{remark}
In fact, writing $O(2)=SO(2) \oplus \mathbb{Z}_2$, with $\mathbb{Z}_2 = \{1,-1\}$ the (closed) subgroups of $O(2)$ fall in three classes \cite{ihrig-golubitsky:84}
\begin{description}
	\item[I] Closed subgroups of $SO(2)$.
	\item[II] Closed subgroups containing $-1$: they are of the form $\Sigma \oplus \mathbb{Z}_2$, $\Sigma$ a subgroup of $SO(2)$.
		\item[III] Closed subgroups of $O(2)$ which are not a subgroup of $SO(2)$ and do not contain $-I$.
\end{description}
In Proposition~\ref{prop:isoG} we considered only the subgroups of $O(2)$ of class I and II. As shown in, e.g., \cite{golubitsky-stewart-etal:88}[Chapter 13, page 131], those subgroups $\Sigma$ are determined by pairs $K \subset H$ of subgroups of $SO(2)$ such that $K$ has index 2 in $H$, see also \cite{olive:19}. If $p: SO(2) \oplus \mathbb{Z}_2 \to SO(2)$, $p(\psi)=det(\psi) \psi$,  is the projection, then $H =p(\Sigma)$ is isomorphic to $\Sigma$ and $\Sigma = K \cup \psi K$ for any $-\psi \in H \backslash K$. It is not difficult to list all such pairs $(K,H)$ using the results in \cite{nganou:12}. None of them produces a subgroup of $E(\mathcal{L}) \times O(2)$ with a nonzero fixed subspace, basically because the color rotations act on $\mathcal{V}$ by multiplications with a magnitude 1 complex exponential.
\end{remark}

	Since $\mathcal{V}$ is $\mathcal{G}$-invariant and $\mathcal{G}$ is compact, we can write $\mathcal{V}$ as a direct sum of $\mathcal{G}$-irreducible subspaces (subspaces such that their only $\mathcal{G}$-invariant subspaces are 0 or the whole subspace), see e.g. \cite{golubitsky-stewart:00}[Theorem 1.22]:
	\[
	\mathcal{V} = \mathcal{V}_1 \oplus \cdots \oplus \mathcal{V}_\rho.
	\]
	It follows easily that
	\[
	{\rm Fix}_{\mathcal{V}}(\Sigma) = {\rm Fix}_{\mathcal{V}_1}(\Sigma) \oplus \cdots \oplus {\rm Fix}_{\mathcal{V}_\rho}(\Sigma).
	\]
	Hence if ${ \rm dim }({\rm Fix}_{\mathcal{V}}(\Sigma)) = 1$ then ${\rm Fix}_{\mathcal{V}}(\Sigma)={\rm Fix}_{\mathcal{V}_j}(\Sigma)$ for some $j$ so that the first step in classifying the planforms associated with a fixed lattice $\mathcal{L}$ is to enumerate each $\mathcal{G}$-irreducible subspace of $\mathcal{V}$. Moreover $\mathcal{V}_j$ is of the form $\mathcal{V}^s_j \times \mathcal{V}^c_j$ where $\mathcal{V}^s_j$ is an eigenspace of $W_s$ irreducible under the action of $E(\mathcal{L})$ and $\mathcal{V}^c_j$ is an eigenspace of $W_c$ irreducible under the action of $O(2)$.

	This has been worked out in the case of $E(\mathcal{L})$ by several authors including \cite{dionne-golubitsky:92}. Dropping the upper index $j$ for simplicity, the irreducible representations of $\mathcal{V}^s$ must be of the form
	\[
	\mathcal{V}^s=\mathcal{V}_{K_1}^s \oplus \cdots \oplus \mathcal{V}_{K_\rho}^s,
	\]
	where  $\mathcal{V}_{K_j}=Vect\left\{z e^{2\pi i \langle K_j, r \rangle} +c.c.,\ z\in\mathbb C \right\}$ , $j=1,\cdots,\rho$. $K_1,\cdots,K_\rho$ are elements of $\mathcal{L}^*$ and the set $\{\pm K_1,\cdots,\pm K_\rho \}$ is an orbit in $\mathcal{L}^*$ of the holohedry $D_4$. As proved in \cite{dionne-golubitsky:92}, there exists only one such representation of dimension 4 and a countable number of them of dimension 8 if we add the further constraint that they have to be translation free. A representation is translation free if there are no (nontrivial) translations in $E(\mathcal{L})$ that act trivially on~\eqref{eq:V}. This requirement ensures that we have found the finest lattice, $\mathcal{L}$, that supports the neutral modes~\eqref{eq:V} \cite{dionne-golubitsky:92}.
	
		The first one corresponds to
	\begin{equation}\label{eq:dim4}
		\left\{
		\begin{array}{ccc}
			K_1 &=& k_1 \\
			K_2 &=& k_2,
		\end{array}
		\right.
	\end{equation}
	the second to
	\begin{equation}\label{eq:dim8}
		\left\{
		\begin{array}{ccc}
			K_1 &=& u k_1+v k_2 \\
			K_2 &=& -v k_1+uk_2 \\
			K_3 &=& v k_1+ u k_2\\
			K_4 &=& -u k_1+ v k_2,
		\end{array}
		\right.
	\end{equation}
	where $u$ and $v$ are relatively prime strictly positive integers such that $u+v$ is odd.

With our notations, we have
		\begin{multline}\label{eq:V}
			\mathcal{V} =\\
			\left\{ u(r,\varrho,\phi)=\sum_{j=1}^k z_j e^{2\pi i \langle K_j,\,r \rangle} e^{2\pi i n \varphi}e^{h,m}(\varrho)+w_je^{-2\pi i \langle K_j,\,r \rangle}e^{2\pi i n \varphi}e^{h,m}(\varrho)+c.c,\,z_j,\,w_j \in \mathbb{C} \right\},\\ k=2,\,4,
	\end{multline}
which is isomorphic to $\mathbb{C}^{2k}$ by $u \to (z,w):=(z_1,\cdots,z_s,\,w_1,\cdots,w_s)$.

If we now consider an isotropy subgroup $\Sigma$ of $\mathcal{H}$, such that ${\rm dim\ Fix}_{\mathcal{V}}(\Sigma)=2$, Proposition~7.2 in Chapter XVI  in \cite{golubitsky-stewart-etal:88} asserts that $\Sigma$ is a twisted group, i.e. there is a pair of subgroups $K \subset G$ of $E(\mathcal{L})$ and a unique homomorphism $\Theta:G \to SO(2)$ such that $\Sigma = G^\Theta:=\{(g,\Theta(g))\in E(\mathcal{L}) \times SO(2),\ g\in G\}.$
		$K$ is  the kernel of $\Theta$: $K=\ker\Theta$. In \cite{dionne-golubitsky-etal:95}[Table 16 on page 157 and Table 21 on page 160] the authors compute all such subgroups and their fixed point subspaces. Using Proposition \ref{prop:isoG} we can obtain the corresponding subgroups of $\mathcal{G}$ and their corresponding fixed point subspaces.

In detail $D_4$ is generated by the $\pi/2$ rotation, noted $R_{\pi/2}^s$ and the reflection, noted $\tau_{r_1}$ through the $r_1$-axis. 
	The action of $E(\mathcal{L})$ on $\mathcal{V}$ induces an action on $\mathbb{C}^{2k}$. The reader will verify that this action is given by
	\begin{equation}\label{eq:actRpi2}
			R_{\pi/2}^s\cdot(z,w) = \left\{
			\begin{array}{cc}
				(z_2, w_1, w_2, z_1 ) & k=2\\
				(z_2, w_1, w_2, z_1,z_4,w_3,w_4,z_3) & k=4
			\end{array}
			\right.,
	\end{equation}
and			
\begin{equation}\label{eq:acttaur1}
			\tau_{r_1}\cdot(z,w)= \left\{
			\begin{array}{cc}
				(z_1, w_2, w_1,  z_2 ) & k=2\\
				(w_4,w_3 , w_2 ,w_1 ,z_4, z_3, z_2, z_1) & k=4
			\end{array}
			\right..
	\end{equation}
Similarly, for the action of $O(2)$, we have
\begin{equation}\label{eq:actphic}
			R_{\varphi_c}^c\cdot(z,w)= 
				e^{2\pi i n \varphi_c}(z,w), \  k=2,\,4	
	\end{equation}
and
\begin{equation}\label{eq:acttauc}
			\tau_c\cdot(z,w)= 
				(\bar{w},\bar{z}), \  k=2,\,4.	
	\end{equation}
	
Writing 
\[
	(z_j,w_j) = \frac 1 2 (z_j+\bar{w}_j, \bar{z}_j+w_j)+\frac 1 2 (z_j-\bar{w}_j, -\bar{z}_j+w_j),\,j=1,\cdots,k,
\]
shows that the projection of $(z,w)$ on $\mathcal{V}_1$ is $\frac{1}{2}(z+\bar{w},\bar{z}+w)$.

The results  in 	\cite{dionne-golubitsky-etal:95} allow us to determine  
all the axial subgroups of $\mathcal{G}$ in dimension 4 (because of \cite{silber-knobloch:91}) and only some of them in dimension 8. They are shown in Table~\ref{table:axialsubgroups} where we have noted $v_d$ the vector $\frac{1}{2}(k_1+k_2)$, $e^s$ the identity of $D_4$ and $e^c$ the identity of $O(2)$.
		\begin{table}[htb]
			\begin{tabular}{lcccc}
				\hline 
				& Generators of $G^{\theta}$ && $p_1\left({\rm Fix}\left(G^{\Theta}\right) \right)$ \\
				\hline 
				&&$\operatorname{dim} \mathcal{V}=4$&\\
				\hline 
$\mathrm{SR}$			& $((-e^s,0),0), ((e^s,\frac{1}{2}),R^c_{\frac 1 2})$ && $z_1=w_1 {\rm real},\,z_2=w_2=0$ \\
$\mathrm{~S} 2$				& $((R_{\pi/2}^s,0), e^c),\, ((\tau_{v_1},0), e^c),\,((e^s,v_d),R_{\frac{1}{2}}^c)$ && \\
				 & $((R_{\pi/2}^s,0), \tau_c),\, ((\tau_{v_1},0), \tau_c),\,((e^s,v_d),R_{\frac{1}{2}}^c\tau_c)$ && $z_{1}=z_{2}=w_{1}=w_{2},\,z_1 {\rm real}$ \\
				
				\hline 
				
				&&$\operatorname{dim} \mathcal{V}=8$ &\\
				
				\hline
$\mathrm{S} 2_{u, v} $				& $((R_{\pi/2}^s,0), e^c),\, ((\tau_{v_1},0), e^c),\,((e^s,v_d),R_{\frac{1}{2}}^c)$ && \\
				
				& $((R_{\pi/2}^s,0),\tau_c),\, ((\tau_{v_1},0),\tau_c),\,((e^s,v_d),R_{\frac{1}{2}}^c\tau_c)$&&$ z_{1}=z_{2}=z_{3}=z_{4}, w=z,\,z_1 {\rm real}$	\\
				
$\mathrm{S} 4_{u, v} $				& $((R_{\pi/2}^s,0),e^c),\, ((\tau_{v_1},v_d),e^c),\,((e^s,v_d),R_{\frac{1}{2}}^c)$&&\\
				& $((R_{\pi/2}^s,0),\tau_c),\, ((\tau_{v_1},v_d),\tau_c),\,((e^s,v_d),R_{\frac{1}{2}}^c\tau_c)$&&$ z_{1}=z_{2}=-z_{3}=-z_{4}, w=z,\,z_1 {\rm real}$\\
				\hline
			\end{tabular}
			\caption{The axial subgroups of $\mathcal{G}$ in dimension 4 and two of them in dimension 8, as well as their corresponding fixed spaces.}
			\label{table:axialsubgroups}
		\end{table}

We can now apply the equivariant branching lemma . In detail we have the following Proposition.

 \begin{proposition}\label{prop:ebl}
 	Let us denote by $(\lambda_n)$ the eigenvalues of $W\in\mathcal L(\mathbb L^2(\Omega))$ and $\gamma_n:=\frac{1}{{\rm Sig}'(-\varepsilon)\lambda_n}$ the bifurcation points.
 	The smooth $\mathcal{G}$-equivariant mapping $F:\mathbb{W}^{m,2}(\Omega)\times\mathbb R\to\mathbb{L}^2(\Omega)$ satisfies the assumptions of the equivariant branching lemma. For each axial subgroup $\Sigma\subset \mathcal{G}$, there exists a unique branch of solutions to $F(V,\gamma_n)=0$ emanating from $(0,\gamma_n)$ where the symmetry of the solution is $\Sigma$. Moreover each of these branches stems from a pitchfork bifurcation.
 \end{proposition}
\begin{proof}
The proof is a direct application of \cite{chossat-lauterbach:00}[Theorem 2.3.2] around $\gamma\approx\gamma_n$. From Section~\ref{sect:stationary} the mapping $F$ is smooth. Its Jacobian at $V=0$ is the linear operator 
\[
dF(0,\gamma_n)=-{\rm Id}+\gamma_n {\rm Sig}'(-\varepsilon) W,
\]
where ${\rm Sig}$ is defined in \eqref{eq:sig}. As $\mathbb W^{m,2}(\Omega)$ is compactly included in $\mathbb L^2(\Omega)$ and $W\in\mathcal L(\mathbb L^2(\Omega))$ is compact, it follows that $W$ is compact from $\mathbb W^{m,2}(\Omega)$ to $\mathbb L^2(\Omega)$. This implies that $dF(0,\gamma_n)$ is a Fredholm operator with index $0$.

From the previous linear analysis, $0$ is an isolated eigenvalue of $dF(0,\gamma_n) \in\mathcal L(\mathbb L^2(\Omega))$ with finite multiplicity. 

Let us denote the reduced equation by $f(v,\gamma)=0,\ v\in\mathcal V$: it is given by the Lyapunov-Schmidt method, see e.g. \cite{golubitsky-schaeffer:84}.
It remains to show that   $d^2_{v\gamma}f(0,0)\neq 0$ to complete the proof of the generic existence/uniqueness of a bifurcating  branch.

We restrict ourselves to the case $\varepsilon=0$ for which $s''(0)=0$. The methods described in e.g.\cite{golubitsky-schaeffer:84}[Chapter VII, page 295], allow us to find, using the fact that $d F$ is hermitian, that $d^2_{v\gamma}f(0,0) = \left\langle \zeta, \partial_\gamma d F(0,0)\cdot\zeta\right\rangle $ where $\zeta\in\mathcal V$ is normalised. A direct computation gives $d^2_{v\gamma}f(0,0)=\frac{1}{\gamma_n}\neq 0$.

The type of bifurcation that occurs can be determined by applying, e.g., \cite{chossat-lauterbach:00}[Theorem 2.3.2]. For each of the two subgroups $G^\theta$ of $\mathcal{G}$ which appear in Table~\ref{table:axialsubgroups} we verify (using the relation $\tau_c R_{\frac 1 2}^c = R_{\frac 1 2} ^c\tau_c$) that the element $g_0:=((e^s,0),R_{\frac 1 2}^c)$ is in the normalizer $N(G^\theta)$ of $G^\theta$ since it satisfies $g_0 g g_0^{-1} =g$ for all generators of $G^\theta$ glven in  Table~\ref{table:axialsubgroups} and acts as $-\mathbf{1}$ on $p_1\left({\rm Fix}(G^\theta)\right)$.
\end{proof}
 This Theorem allows us to discover a specific class of planforms, i.e. some of those satisfying \eqref{eq:fix1}. In general and generically there may exist solutions such that ${ \rm dim }({\rm Fix}_{\mathcal{V}}(\Sigma)) > 1$. Our approach here will not allow us to find these. Note however that the assumption \eqref{eq:fix1} is the most commonly made. Exceptions to this can be found in \cite{busse:75,chossat:83,chossat-lauterbach-etal:90}.

\begin{figure}[htb]
\centerline{
\includegraphics[width=0.6\textwidth]{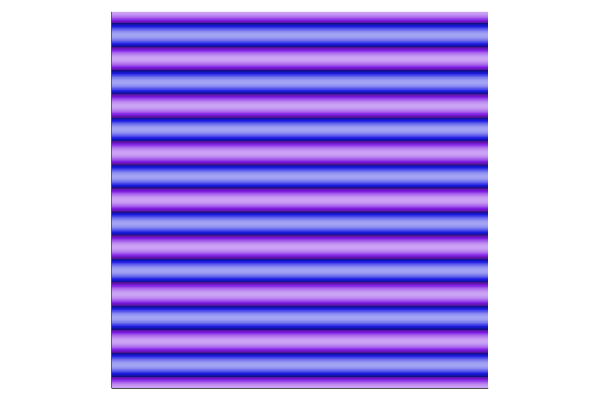}
\hspace{0.01cm}
\includegraphics[width=0.6\textwidth]{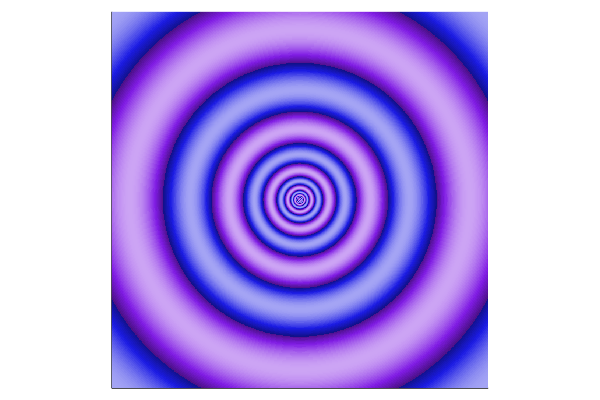}
}
\caption{An example of a planform whose equation is given by \eqref{eq:SR}: Left in cortical coordinates, Right in retinal coordinates.}
\label{fig:planform-4D-SR}
\end{figure}

\begin{figure}[htb]
\centerline{
\includegraphics[width=0.6\textwidth]{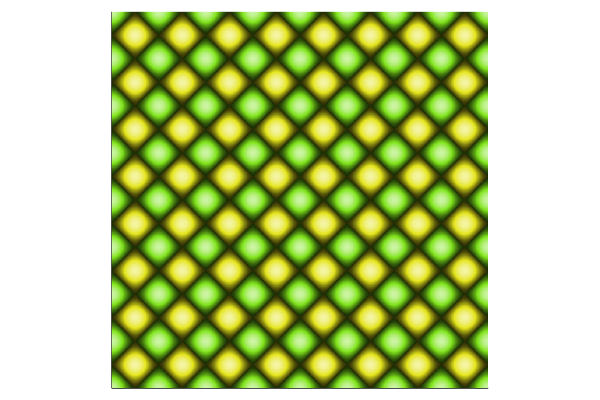}
\hspace{0.01cm}
\includegraphics[width=0.6\textwidth]{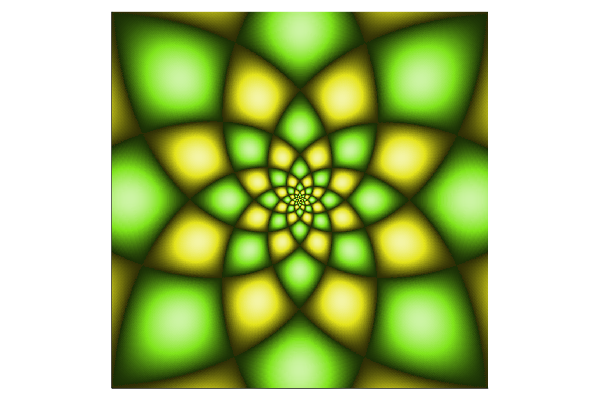}
}
\caption{An example of a planform whose equation is given by \eqref{eq:S2}: Left in cortical coordinates, Right in retinal coordinates.}
\label{fig:planform-4D-S2}
\end{figure}

\section{Examples of planforms}\label{sect:explanf}
Using Table~\ref{table:axialsubgroups} and \eqref{eq:V}, it is easy to write down the analytical expressions of the four types of planforms. In the case where ${\rm dim}\mathcal{V}=4$, we have
\begin{equation}\label{eq:SR}
SR(r,\varrho,\varphi)= \cos2\pi r_1 e^{h,m}(\varrho)\cos (2\pi n \varphi),
\end{equation}
and
\begin{equation}\label{eq:S2}
S2(r,\varrho,\varphi) = \left( \cos 2\pi r_1+\cos 2\pi r_2 \right)  e^{h,m}(\varrho)\cos (2\pi n \varphi).
\end{equation}
The first type of planform is called standing rolls (SR) or stripes, the second type is called spots. 

We show in Figure~\ref{fig:planform-4D-SR} an example of standing rolls (or stripes) described by equation~\eqref{eq:SR} in cortical and  retinal coordinates, respectively. The relation between cortical and retinal coordinates is discussed in Remark~\ref{rem:logpolar}. 
Similarly we show in Figure~\ref{fig:planform-4D-S2} an  example of spots described by equation~\eqref{eq:S2}. 

\begin{figure}[h]
\centerline{
\includegraphics[width=0.6\textwidth]{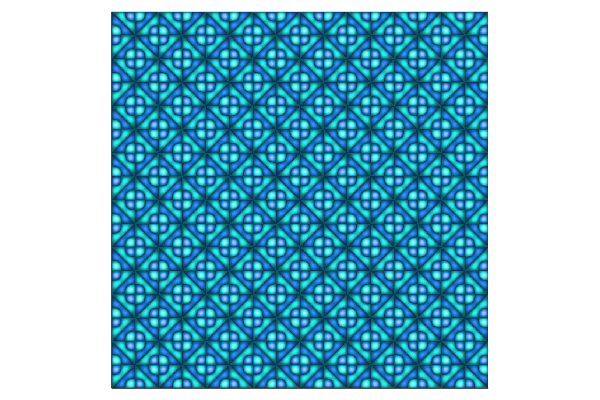}
\hspace{0.01cm}
\includegraphics[width=0.6\textwidth]{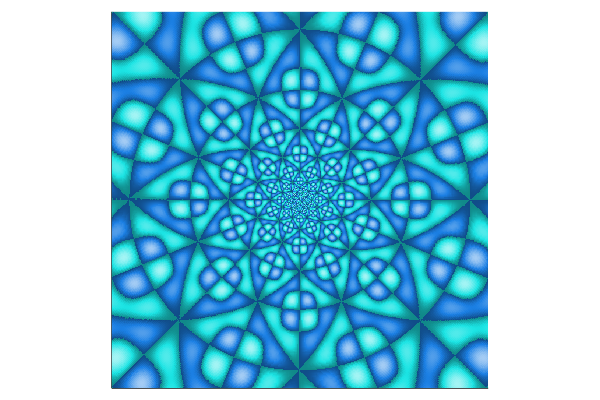}
}
\caption{An example of a planform whose equation is \eqref{eq:S4uv}: Left in cortical coordinates, Right in retinal coordinates.}
\label{fig:planform-8D-2}
\end{figure}

In the case ${\rm dim}\mathcal{V}=8$ we have the two families of planforms
\begin{multline}
S2_{u,v}(r,\varrho,\varphi) =\Big(\cos 2\pi(u r_1+v r_2)+\cos 2\pi(-v r_1+u r_2)+\\
\cos 2\pi(v r_1+u r_2)+\cos 2\pi(-u r_1+v r_2) \Big)   e^{h,m}(\varrho) \cos (2\pi n \varphi)\label{eq:S2uv},\end{multline}
\begin{figure}[h]
\centerline{
\includegraphics[width=0.6\textwidth]{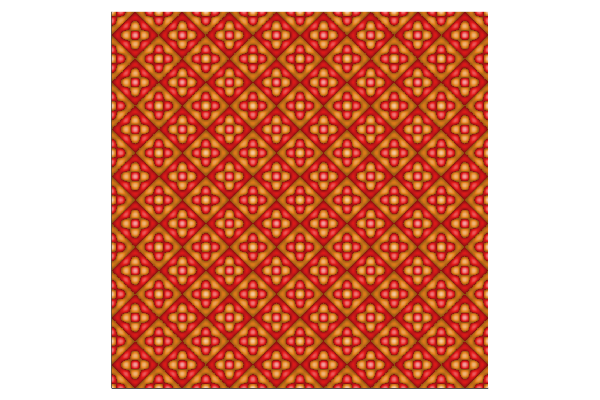}
\hspace{0.01cm}
\includegraphics[width=0.6\textwidth]{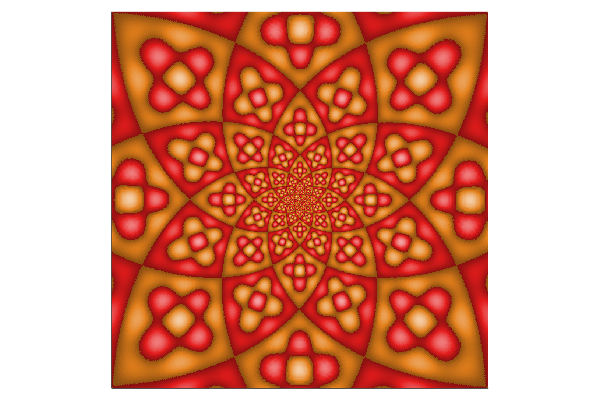}
}
\caption{An example of a planform whose equation is \eqref{eq:S2uv}: Left in cortical coordinates, Right in retinal coordinates.}
\label{fig:planform-8D-1}
\end{figure}

and
\begin{multline}
S4_{u,v}(r,\varrho,\varphi) =\Big(\cos 2\pi(u r_1+v r_2)+\cos 2\pi(-v r_1+u r_2)-\\
\cos 2\pi(v r_1+u r_2)-\cos 2\pi(-u r_1+v r_2) \Big)  e^{h,m}(\varrho) \cos (2\pi n \varphi) \label{eq:S4uv},\end{multline}
where $u$ and $v$ are relatively prime strictly positive integers such that $u+v$ is odd.
Figure~\ref{fig:planform-8D-1} shows an example of a planform $S2_{1,2}$ and 
Figure~\ref{fig:planform-8D-2} shows an example of a planform $S4_{3,2}$.
\begin{remark}\label{rem:colorrep}
To determine the color represented at cortical location $(r_1,r_2)$, and produce the images in  Figures~\ref{fig:planform-4D-SR}-\ref{fig:planform-8D-2}, we compute $c_{\rm max}(r_1,r_2) = {\rm argmax}_{\varrho,\varphi}\Upsilon\left(V^{\rm eq}(r_1,r_2,\varrho,\varphi)\right)\in \Omega_c$  (corresponding to the maximum activity of the neural mass with spatial coordinates $(r_1,r_2)$), $V^{\rm eq} \in \{SR,\,S2,\,S2_{u,v},\,S4_{u,v}\}$,  and define the Luminance  to be the corresponding activity level $\Upsilon\left(V^{\rm eq}(r_1,r_2,c_{\rm max}(r_1,r_2))\right)$.   In case there are several argmax, we arbitrarily select one of them. At each cortical location $(r_1,r_2)$ we display the color using the  HSL color coordinates, see Appendix~\ref{section:color space} and \cite{joblove-greenberg:78}:
\begin{align*}
H &= 2\pi \varphi\\
S &= \sqrt{\varrho}\\
L &= \Upsilon\left(V^{\rm eq}(r_1,r_2,c_{\rm max}(r_1,r_2))\right)
\end{align*}
\end{remark}
\begin{remark}
In Figures~\ref{fig:planform-4D-SR}-\ref{fig:planform-8D-2} the value of $n$ in Proposition~\ref{prop:eigenwha}  is equal to 6. The value of $\xi_c$ in Proposition~\ref{prop:eigenwhm} is equal to 2 and $x_n$ is the fourth strictly positive root of equation \eqref{eq:xwhm}. This implies that the product $e^{h,m}(\varrho) \cos ( 2\pi n \varphi)$ has several minima and maxima so that there are several values of $(\varrho,\varphi)$ where the maximum of  $\Upsilon(SR(r,\varrho,\varphi))$(respectively of $\Upsilon(S2(r,\varrho,\varphi))$, $\Upsilon(S2_{u,v}(r,\varrho,\varphi))$ or $\Upsilon(S4_{u,v}(r,\varrho,\varphi))$) with respect to $(\varrho,\varphi)$ is reached. As  in Section~\ref{subsec:colorequilib} we have chosen arbitrarily one of them.
\end{remark}

	\section{Numerical bifurcation analysis}\label{sec:hallucinations} 
	
	We recall that we are interested in visual hallucinations which are stable solutions of \eqref{eq:wcc} and exist in fairly large regions of the parameter space.
	The equivariant branching lemma (EBL) provided a set of stationary solutions of \eqref{eq:wcc} for a constant external current $I_{\rm ext} = 0$. It relies on the Lyapunov-Schmidt reduction (or more generally on the existence of a center manifold) which is local by essence, \textit{i.e.} valid in a given neighborhood of $(0,\gamma_n)$ in $\mathcal F\times\mathbb R$. The size of this neighborhood quantifies the predictability of the local theory and is bounded by $\min(|\gamma_{n+1}-\gamma_n|,|\gamma_{n-1}-\gamma_n|)$. Hence, for large cortices, it is vanishingly small. 
	
	We thus use numerical bifurcation analysis to assess our theoretical predictions beyond their above domain of validity. We present some numerical results concerning the equilibria of \eqref{eq:wcc} for a constant external current $I_{\rm ext} = 0$  as functions of different parameters. Because of the volume of data required to explore the full four-dimensional model (2 dimensions in space and 2 in color) we restrict ourselves to a one-dimensional color space by considering a diameter of the unit disc determined by its hue angle $\varphi_0$ (between 0 and 1 to be consistent with our notations). A point on this diameter is characterized by its polar coordinates $(2 \pi \varphi_0, \rho)$ or $(2 \pi (\varphi_0 + \frac{1}{2} \mod 1),\rho)$, $\rho \in [0,1]$. The relevant color coordinate is therefore $c:=\pm \rho \in [-1,\,1]$. 
	
Note that the problem associated has now the symmetry group $(D_4\overset{\cdot}{+}T_2) \times \mathbb{Z}_2$ instead of $(D_4\overset{\cdot}{+}T_2) \times O(2)$. This has been extensively studied \cite{silber-knobloch:88,dionne-silber-etal:97}. A key remark is that the additional reflection symmetry has no effect on the bifurcation problems associated with the square lattice\footnote{This was pointed out to us by Pascal Chossat.}.

	In the first bifurcation diagram (Section~\ref{subsect:firstbif}), we observe that the type of the bifurcated branches and their stability predicted by the EBL are valid outside the neighborhood for the first bifurcation point, of dimension 4. The second bifurcation point, of dimension 8, however provides an example of a branch which becomes stable. Additionally, all stable patterns are alike (stripes) and in agreement with the EBL predictions.
	
	In the second bifurcation diagram (section~\ref{subsect:secondbif}), we switched the bifurcation points, forcing the first one to be of dimension 8,  in hope to see new stationary states, not like stripes. As before, only the stripes are stable and the EBL predictive power, outside the neighborhood, is not bad.
	
	In search for more interesting hallucinations and based on \cite{veltz-faugeras:10}, we changed the criticality of the bifurcation points in the third bifurcation diagram (section~\ref{subsect:normalform}) hoping to stabilize the patterns with the creation of fold bifurcations. The EBL predictions concerning the stability of the branches are challenged very quickly. In passing, this provides an example of stable spots. Additionally, the bistability between the state $V=0$ and the branch of stripes suggests snaking branches. These snaking branches are very interesting because they give birth to stable spatially localized visual hallucinations. The last bifurcation diagram shown in Section~\ref{subsect:fourthbif} is dedicated to finding snaking branches.
	
	To conclude, it seems that we can rely on the EBL to predict the existence of patterns that survive in large parameter domains but not to predict their stability. From the simulations, it seems that the spatial structure of the solutions does not vary much along the branches and this suggests the possibility to extend the results from \cite{rabinowitz:71} to the present setting. If such result were true, it would imply that the EBL is a very valuable tool at elucidating the spatial structure of the visual hallucinations.
		
	\subsection{Color representation of the equilibria of \eqref{eq:wcc}}\label{subsec:colorequilib}
\begin{figure}
		\centering
		\includegraphics[width=0.95\linewidth]{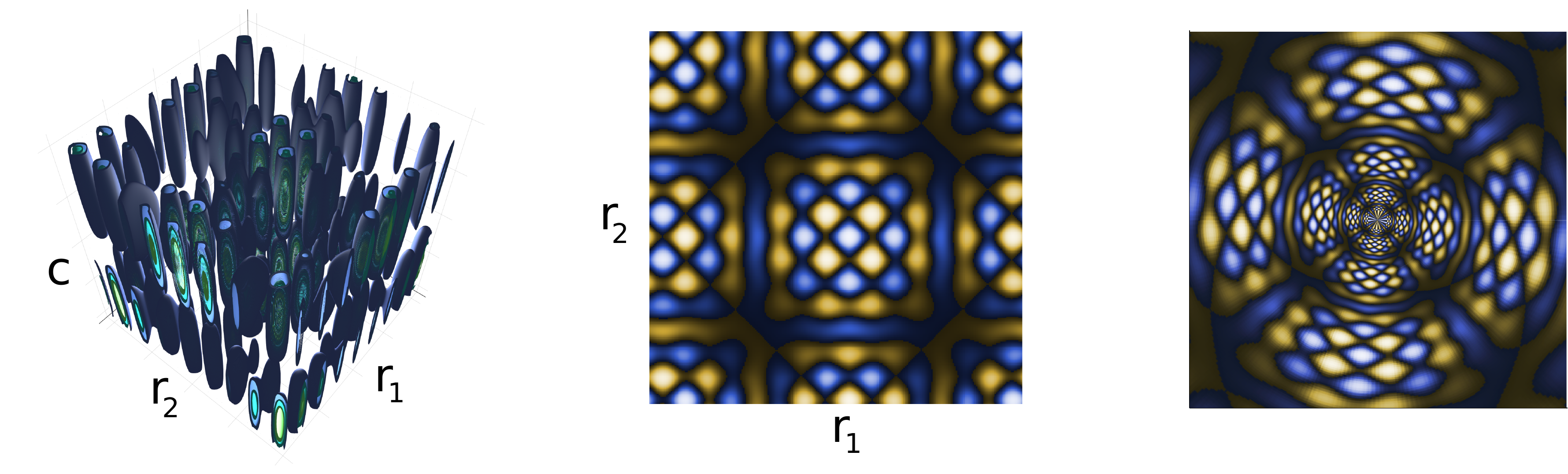}
		\caption{Left: example of level sets of an equilibrium, a stationary solution, of \eqref{eq:wcc}. Middle: Color image representation of the equilibrium shown on the left. Right: Same as Middle but in retinal coordinates.}
		\label{fig:patternmarronbd3}
	\end{figure}

	 The equilibria of \eqref{eq:wcc} are represented by a color image in a way similar to what is described in Section~\ref{sect:explanf}. In Figure~\ref{fig:patternmarronbd3}-Left, we show an example of such an equilibrium $V^{\rm eq}(r_1,r_2,c)$ where we plot some of its level sets  in the three-dimensional space of coordinates $(r_1,r_2,c)$.
	To determine the color represented at cortical location $(r_1,r_2)$, and produce the image in  Figure~\ref{fig:patternmarronbd3}-Middle, we compute $c(r_1,r_2) = {\rm argmax}_c\Upsilon\left(V^{\rm eq}(r_1,r_2,c)\right)\in[-1,1]$  (corresponding to the maximum activity of the neural mass with spatial coordinates $(r_1,r_2)$)  and define the Luminance  to be the corresponding activity level $a(r_1,r_2) := \Upsilon\left(V^{\rm eq}(r_1,r_2,c(r_1,r_2))\right)$.   In case there are several argmax, we arbitrarily select one of them as in Section~\ref{sect:explanf}.  At each cortical location $(r_1,r_2)$ we display the color using the  HSL color space, see Appendix~\ref{section:color space} and \cite{joblove-greenberg:78}:
	
	\begin{equation}\label{eq:color}
	\left\{
	\begin{array}{lcl}
	H & = & 2\pi \varphi_0+\frac \pi 2 \times(1-{\rm sign}(c(r_1,r_2)))\\
	S & = & |c(r_1,r_2)|\\
	L & = & a(r_1,r_2)
		\end{array}
		\right.
	\end{equation}
	where
	\[
	{\rm sign}(x) = \left\{
	\begin{array}{ll}
		+1 & \text{if } x \geq 0\\
		-1 & \text{otherwise}
	\end{array}
	\right.
	\]
	and obtain the result shown in Figure~\ref{fig:patternmarronbd3}-Middle. $\varphi_0=\frac 1 8$ is defined in Appendix~\ref{section:color space}.
	\begin{remark}
		Note that this way of displaying our results is a severe simplification of our model which says that at every location $(r_1,r_2)$ in the cortex, color is represented by the \emph{function} $c \to \Upsilon(V^{eq}(r_1,r_2,c))$ and not by the three \emph{ numbers} \eqref{eq:color}, see \cite{koenderink:10} for an interesting discussion of these issues and much more. 
	\end{remark}
	
	\begin{remark}\label{rem:logpolar} The visual cortex of several species (like monkeys) has the property of being retinotopically organized, e.g. \cite{schwartz:77,petitot:17}. That is, there is a one-to-one mapping from retinal coordinates to cortical ones, the mapping in humans being approximately log polar. Thus, we need to apply an inverse log polar transform to the equilibrium shown in Figure~\ref{fig:patternmarronbd3}-Middle in V1 coordinates to obtain the result shown in Figure~\ref{fig:patternmarronbd3}-Right in retinal coordinates.
	\end{remark}

	\subsection{Connectivity functions}
\begin{figure}[tbph!]
		\centering
		\includegraphics[width=0.95\textwidth]{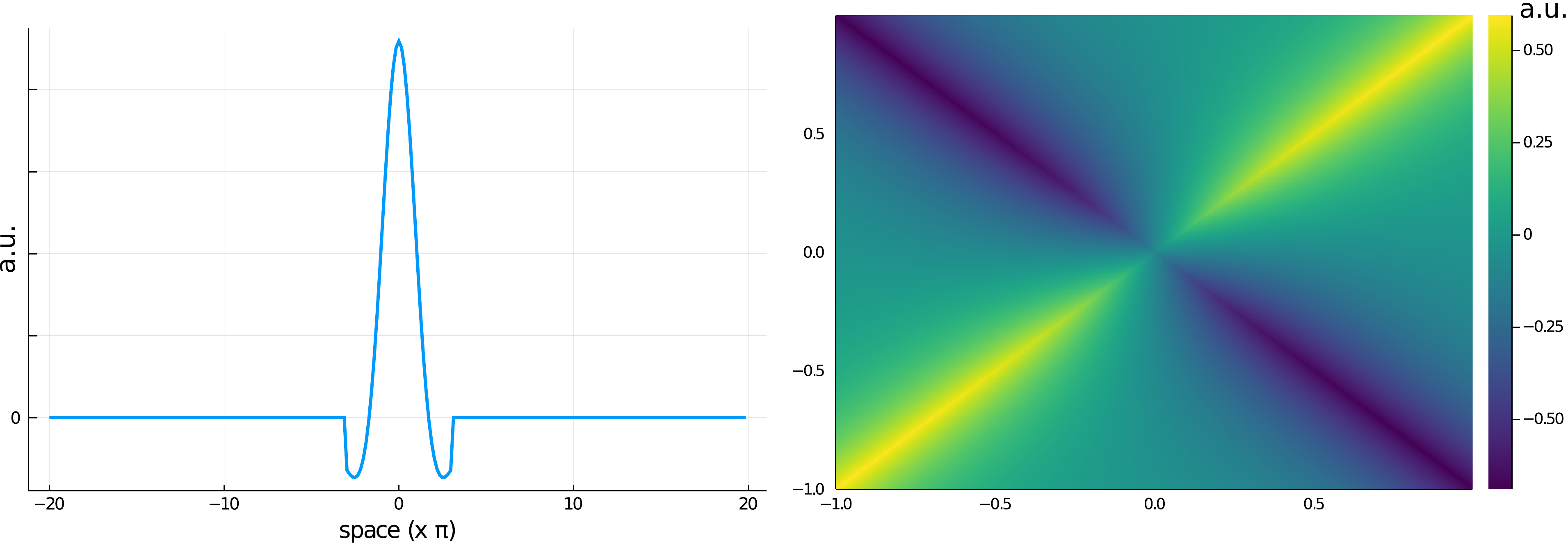}
			\caption{Left: Typical shape of the spatial connectivity $w_s$ used in the numerical experiments. Right:  Heatmap plot of the color connectivity function $w_c(c,c')$ used in the numerical experiments.}
		\label{fig:jspatial-bif1}
	\end{figure}

	In all subsequent examples, we use the following spatial connectivity function $w_s(r)$ of the form \eqref{eq:g} whose parameters $\mu_s$, $\nu_s$ and $\beta_s$ are chosen so that it satisfies \eqref{eq:balanced}:
\begin{equation}\label{eq:wsnumer}
w_s(r)=A\left(\mu_s e^{-\|r\|_{2}^{2} / 2 \pi^{2}}-\nu_s e^{-\|r\|_{2}^{2} / 2 (\beta_s\pi)^{2}}\right)\mathbf 1_{\|r\|\leq B}.
\end{equation}
	The specific values of these parameters corresponding to different numerical experiments are shown in the captions of  Figures~\ref{fig:bdiag1}-~\ref{fig:bdiag3a}.
	A cross-section is shown in Figure~\ref{fig:jspatial-bif1}-Left: it has the traditional Mexican hat shape.

	Similarly $w_c(c,c')$ is defined by
	\[
	w_c(c,c') := \mu_c e^{-\alpha_c |c-c'|} - \nu_c e^{-\beta_c |c+c'|}
	\] 
	with the following values of the parameters:
	\begin{align*}
		\alpha_c = 0.3\quad & \beta_c =0.4\\
		\mu_c  = 0.6,\quad & \nu_c = 0.69.
	\end{align*}	
	A heatmap of this function is shown in Figure~\ref{fig:jspatial-bif1}-Right.
	
	\subsection{Numerical experiments}\label{section:methods}

	All numerical computations were performed in the \href{https://julialang.org}{Julia} programming language (version 1.4.2). The bifurcation diagrams were computed using a pseudo-arclength continuation method  implemented in the package \textit{BifurcationKit.jl}\cite{veltz:20} with version v0.1.2	. The continuation is based on a Newton–Krylov method to solve \eqref{eq:wccs} with GMRES linear solver \cite{barret-berry-etal:94}. The computation of the eigen-elements, to detect the bifurcation points, is based on the Arnoldi algorithm \cite{saad:11}. The linear and eigen solvers are both implemented in the package \href{https://github.com/Jutho/KrylovKit.jl}{\textit{KrylovKit.jl}}. The nonlinear equations \eqref{eq:wccs} were solved at tolerance $10^{-11}$ in the supremum norm. The bifurcation points were located using a bisection algorithm (on the number of unstable eigenvalues) leading to a precision of $10^{-4}$ on the value of the parameter $\gamma$, see \eqref{eq:sig}.
	
	Let us describe how \eqref{eq:F} is implemented. The connectivity kernels were computed with the three-dimensional Fast Fourier Transform (3D FFT) on Graphics Processing Units (GPU) based on the package \href{https://github.com/JuliaGPU/CUDA.jl}{\textit{CUDA.jl}} (see \cite{besard-foket-etal:18, besard-churavy-etal:19}).

	In order to compute the bifurcating branches, the reduced equations (see \cite{golubitsky-stewart-etal:88} or proof of proposition~\ref{prop:ebl}) at the bifurcation point were computed thereby yielding a system of polynomial equations of degree 3 in a number of variables equal to the dimension of the kernel of the Jacobian. The roots of these polynomial equations are then computed and used as guesses for points on the bifurcated branches which are corrected using a Deflated-Krylov-Newton (see \cite{farrell-birkisson-etal:15}) to prevent converging to the trivial solution. This provides an entirely automatic procedure to find the bifurcated branches at a bifurcation point of any dimension. We call this procedure \textit{automatic branch switching} (aBS).
	
	The whole program runs \textbf{entirely} on GPU, a V100 Nvidia card with 32Gb of RAM. The computations are next to impossible to run without a GPU albeit perhaps on a cluster. In the experiments, we use the values $N_{r_1}= N_{r_2}=256, N_c=64$. Using a finer discretisation would have helped but we were limited by memory when computing eigen elements. Indeed, computing the branches without stability is a matter of a few minutes. However, we found necessary to use a Krylov Space of size $\approx 100$, due to the symmetries, in order to compute the eigenvalues, and this limited the size  $N_{r_1}\times N_{r_2}\times N_c$ of the discretisation.  
	
	\subsection{First bifurcation diagram}\label{subsect:firstbif}
	
	\begin{figure}[htbp!]
		\includegraphics[width=0.75\textwidth]{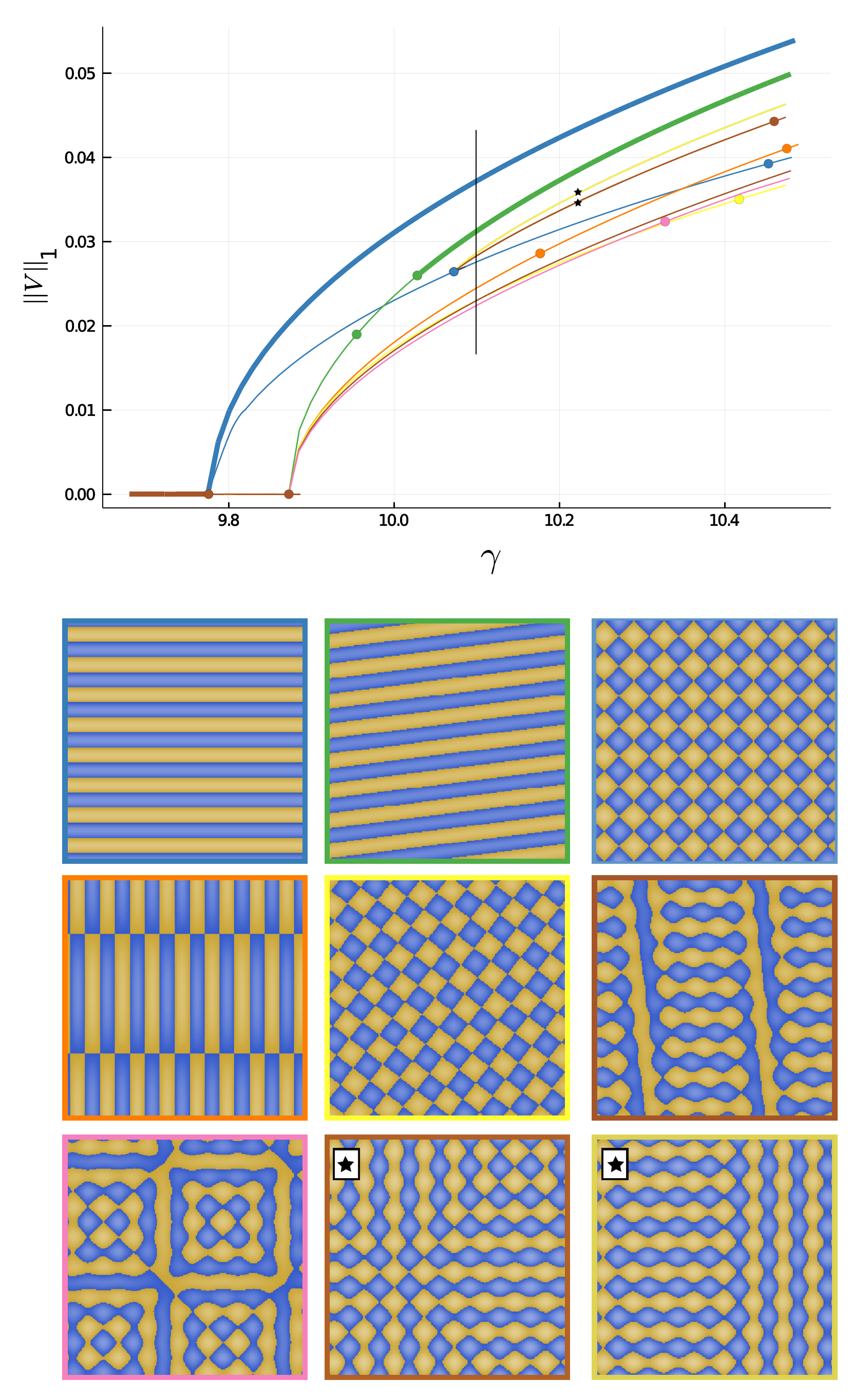}  
		\caption{Parameters $\varepsilon =0,\, \Omega_s = (-20\pi, 20\pi)^2,\, \Omega_c = (-1,1), \, \beta_s = 1.9,\, B = 3\pi,\, A = 1.0$. 
			Equilibria as functions of the nonlinear gain $\gamma$. The stable parts of the branches are indicated with thick lines. The dots indicate bifurcation points. The discretization is $256\times 256\times 64$. Only the first two primary bifurcation points are shown. The colours around the images are those of the corresponding branches. The first seven images represent the intersections of the vertical black line with the bifurcated branches. The last two images correspond to the two black stars in the bifurcation diagram.} 
		\label{fig:bdiag1}
	\end{figure}
	
	Figure~\ref{fig:bdiag1} represents the variation of the equilibria (represented by their $\mathbb{L}^1$ norm) as functions of the slope  $\gamma$ of the activation function $s$, see \eqref{eq:sig}. The trivial equilibrium $\bar{V}=0$ loses stability at a first bifurcation point around $\gamma \approx 9.78$. 
	The dimension of the first primary bifurcation point is four whereas the dimension of the second one is eight. Given the symmetries of the network, it is straightforward to conclude that the first primary bifurcation is a Pitchfork with $D_4$ symmetry group. 
	
	Using the aBS procedure (see Section~\ref{section:methods}), we computed the different bifurcating branches from the first two primary bifurcation points. We know that two branches (at most) bifurcate from the first point: one branch of stable stripes (the blue thick line)  and one branch of unstable spots (the blue thin line).  At the 8D primary bifurcation point, we found four  branches with patterns in agreement with the ones described in \cite{dionne-silber-etal:97}.
	
	We also computed the bifurcated branches (shown in yellow and brown) from the secondary bifurcation point on the spot branch (in light blue) hoping to find new stable patterns.
	
	This diagram hints at the fact that only the stripes are stable. 	
	At the bottom of Figure~\ref{fig:bdiag1} we display the images corresponding to some of the equilibria shown in the plot at the top of the Figure. The images are embedded in a frame the same color as the curve on which the corresponding equilibrium sits. A black star has been added to the two images corresponding to the two equilibria (also marked with a  black star) sitting on the line of equilibria branching out of the line of unstable spots (thin light blue). 
	
	\subsection{Second bifurcation diagram}\label{subsect:secondbif}
	
	\begin{figure}[htbp!]
		\includegraphics[width=0.75\textwidth]{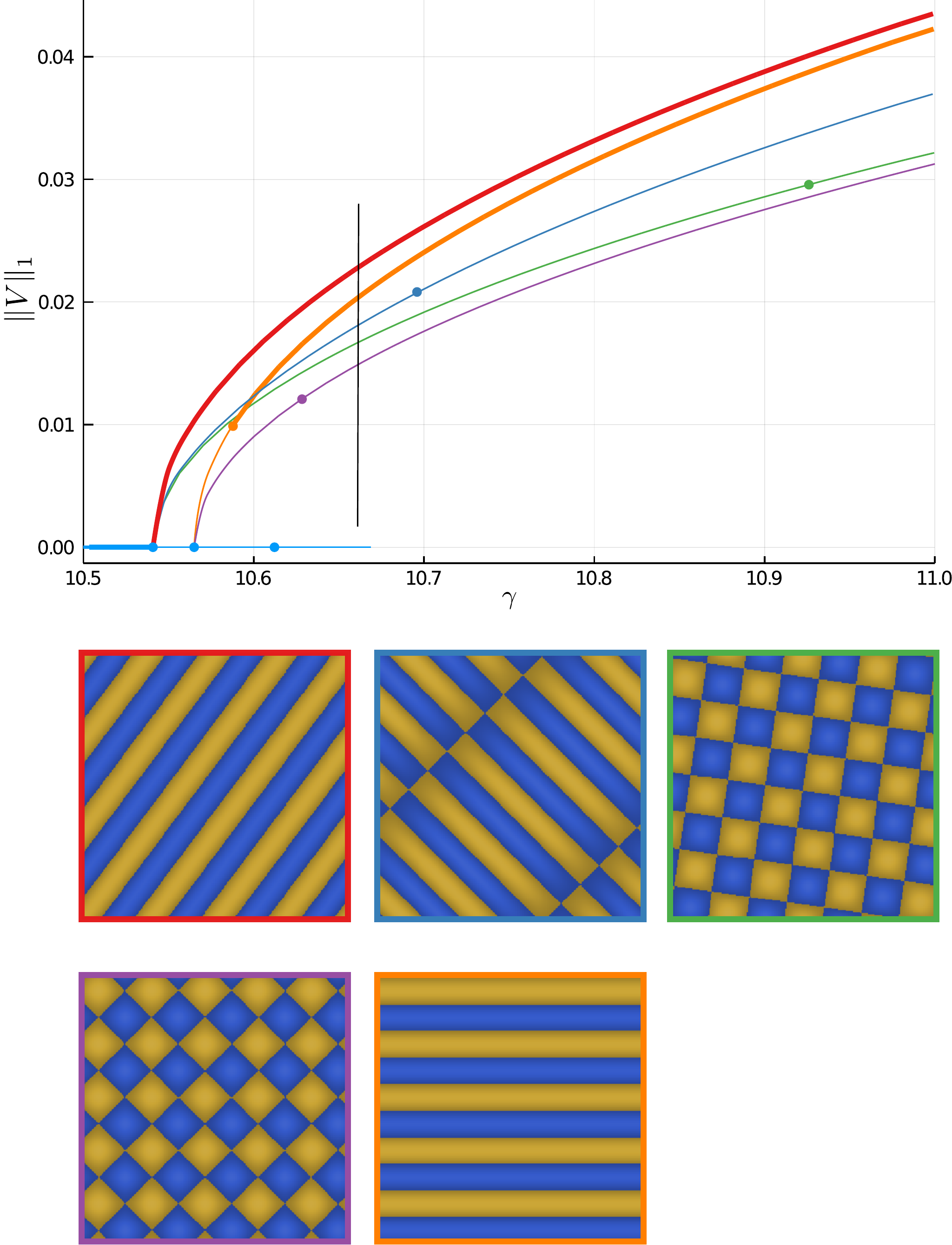}
		\caption{$\varepsilon =0.5,\, \Omega_s =(-20\pi,20\pi)^2,\, \Omega_c = (-1,1),\,  \beta_s = 20.9,\, B = 6\pi,\, A = 1.0.$ Equilibria are shown as functions of the nonlinear gain $\gamma$. The stable parts of the branches are indicated with thick lines. The dots indicate bifurcation points. Only the first three primary bifurcation points are shown. As in Figure~\ref{fig:bdiag1} the colours surrounding the images correspond to those of the branches found at the intersections with the vertical black line. }
		\label{fig:bdiag2}
	\end{figure}
	
	Next, we modify a bit the connectivity to make the dimension of the first bifurcation point 8 and that of the second one 4. We did this hoping to find more interesting stable patterns than in the previous diagram. Using aBS (see Section~\ref{section:methods}), we compute the different bifurcating branches from the bifurcation points. Again, only the stripes are found to be stable. The bifurcated patterns from the 8D bifurcation point are in agreement with the ones referenced in \cite{dionne-silber-etal:97}. In fact, the first two bifurcation diagrams collectively show all the possible patterns that can bifurcate from an 8D bifurcation point.
	
	\subsection{Third bifurcation diagram, changing the criticality}\label{subsect:normalform}
	
	\begin{figure}[htbp!]
		\includegraphics[width=0.8\textwidth]{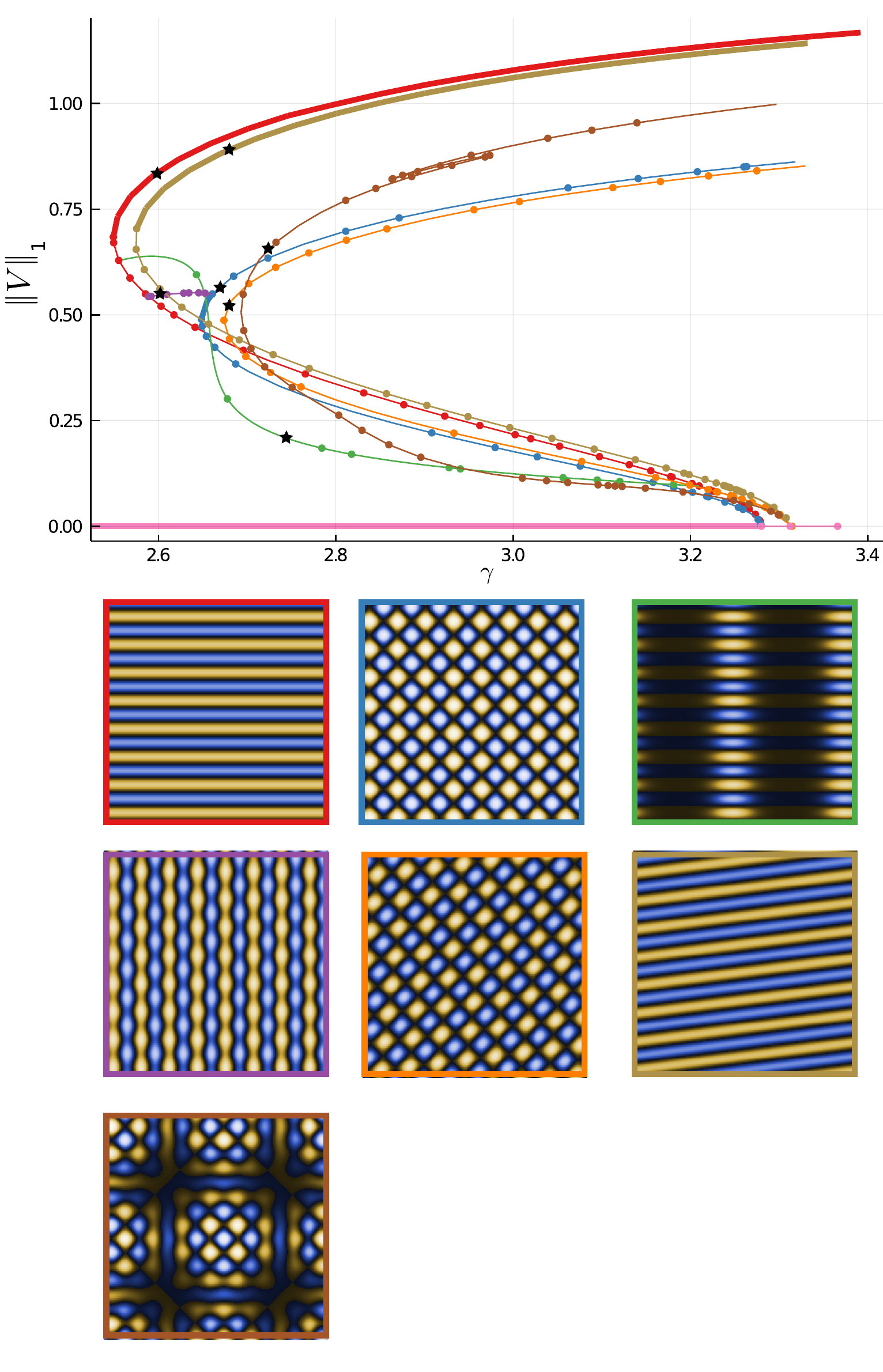}
		\caption{$\varepsilon = 2.3,\, \Omega_s =(-20\pi ,20\pi)^2,\, \Omega_c = (-1,1),\, \beta_s = 1.9,\, B = 3\pi,\,  A = 9.$ Equilibria as function of the nonlinear gain $\gamma$. The stable parts of the branches are indicated with thick lines. The dots indicate bifurcation points. Only the first three primary bifurcation points are shown. The colour surrounding the image is that of its corresponding branch. Their location on the bifurcation diagram is shown with a black star.}
		\label{fig:bdiag3}
	\end{figure}

	We then change the criticality of the bifurcation points  hoping to induce new stable equilibria. To this end, we use the following observation. The normal form of the 4D bifurcation point is \cite{hoyle:06}
	
	\begin{equation*}
		\left\{\begin{array}{l}
			\dot{z}_{1}=z_{1}\left(\alpha+\beta_1\left|z_{1}\right|^{2}+\beta_2\left|z_{2}\right|^{2}\right) \\
			\dot{z}_{2}=z_{2}\left(\alpha+\beta_1\left|z_{2}\right|^{2}+\beta_2\left|z_{1}\right|^{2}\right)
		\end{array}\right. 
	\end{equation*}
	where $z_i\in\mathbb C,\alpha,\beta_i\in\mathbb R$, $i=1,\,2$, and (for example for $\beta_1$, see \cite{veltz:11b}  )
	\[
	\beta _1/ \gamma_0^{3} \lambda_0=\gamma_1 s_{2}^{2}A+s_{3} / 2
	\]
	where $A\in\mathbb R, s_2=Sig^{(2)}(\varepsilon),s_3=Sig^{(3)}(\varepsilon)$, $\gamma_0$ is the value of $\gamma$ at the bifurcation point and $\lambda_0$ is the associated eigenvalue of  $W$. We recall that for a threshold $\varepsilon=0$, $s_2=0, s_3 < 0$ and thus the Pitchfork is supercritical. When varying the threshold $\varepsilon$, the ratio $\frac{s_3}{s_2^2}$ takes on any value, thus we can change the criticality by altering the threshold. Compared to the previous diagram, we use $\varepsilon=2.5$ and scale the connectivity to restrict the value of $\gamma$ at the first bifurcation point.
	
	Using this procedure, we obtain the bifurcation diagram shown in Figure~\ref{fig:bdiag3}. We observe that all primary branches bifurcate sub-critically and thus present a saddle-node bifurcation as proved in \cite{veltz-faugeras:10}.
	The first two bifurcation points are 4D and 8D, respectively. The stripes branches (red or light brown) from either bifurcation points become stable after the saddle-node bifurcation. Remarkably, the spots branch (blue) from the first 4D bifurcation point is stable close to its saddle-node bifurcation. Using aBS, we find a connection (in violet) between the spot branch and the stripe branch.
	
	A very similar situation was found in \cite{uecker-wetzel:14} for the Selkov-Schnakenberg model. The fact that there is bistability between the trivial solution and the stripes (or spots) in a gradient system such as \eqref{eq:wcc} (see \cite{veltz:11b} for a proof) opens the doors for the existence of localized patterns and in particular snaking (see \cite{knobloch:15} for a review) of localized solutions.
	
	Snaking is interesting in the context of visual hallucinations because it produces stable patterns. These snaking branches usually originate from the bifurcation points on the stripes (resp. spots) branch. We thus computed the secondary branches on the spots branch and show one of them (in green). The green branch connects two bifurcation points on the stripes branch. It consists of localized stripes as seen in Figure~\ref{fig:bdiag3}. Unfortunately, it is neither stable nor does it feature snaking. We come back to this point in the next section. Finally, we also computed a secondary branch (brown) from the branch of spots (primary branch of the 8D bifurcation point). It shows another type of (unstable) localized patterns of spots.

	\subsection{Fourth bifurcation diagram, snaking branch of localized solutions}\label{subsect:fourthbif}
	In this last example, we focus on finding snaking branches for \eqref{eq:wcc}. Theoretical analysis (one of the earliest is \cite{chapman-kozyreff:09})  shows that snaking occurs in a region  which  is exponentially small in $\epsilon = \gamma_1 - \gamma_1^{SN}$ where $\gamma_1$ is the value of $\gamma$ at the 4D bifurcation point and $\gamma_1^{SN}$ is that at the location of the saddle-node bifurcation point on the branch of stripes.  In \cite{uecker-wetzel:14} the authors study this behaviour numerically. Thus, to observe snaking, we have to increase $\varepsilon$. By performing codim 2 continuation of the saddle-node bifurcation and the first primary bifurcation point in the $(\gamma, \varepsilon)$ plane, we were able to select a better value of the threshold $\varepsilon$ than in Figure~\ref{fig:bdiag3}.
	
\begin{figure}[htbp!]
		\includegraphics[width=0.965\textwidth]{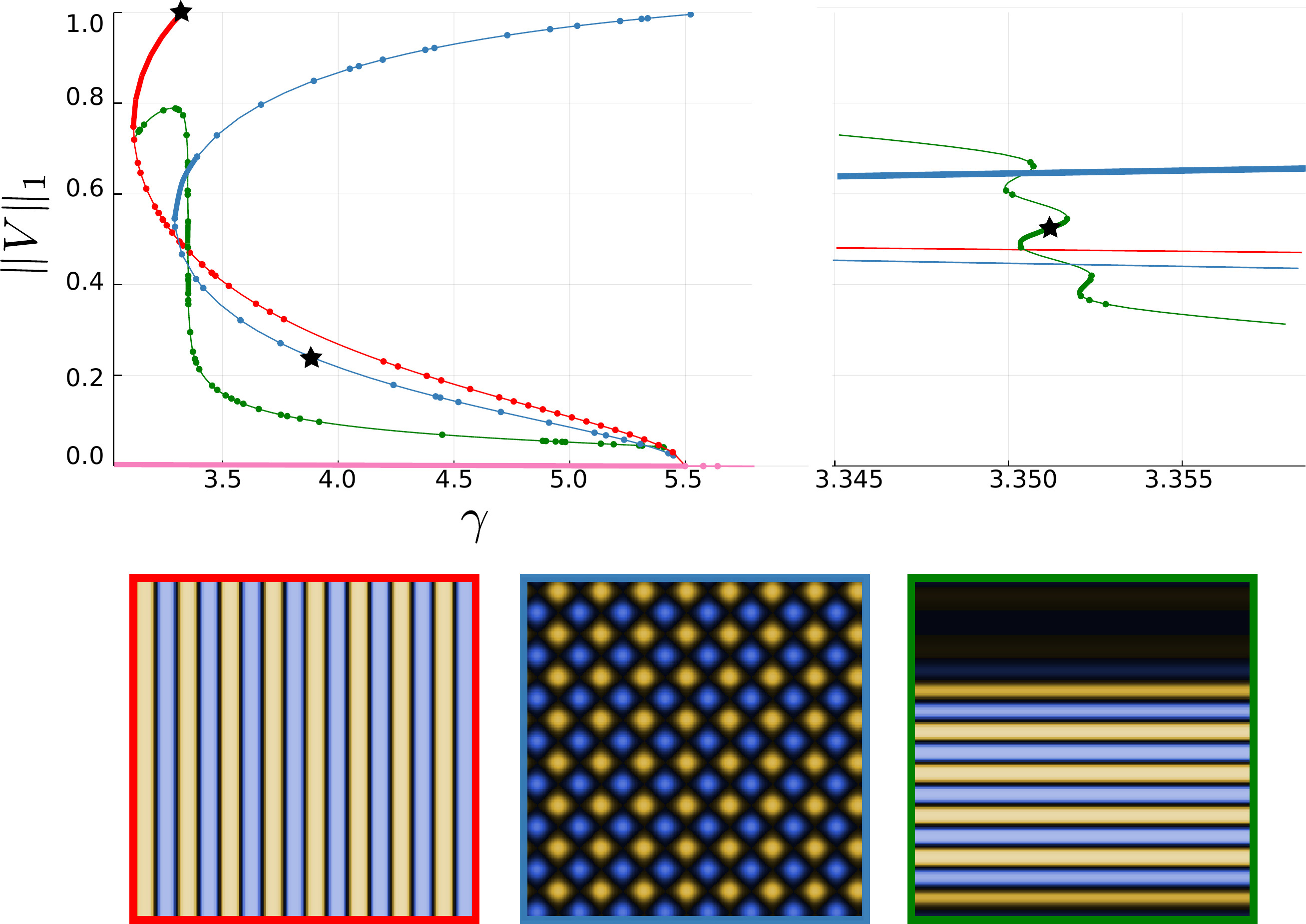}
		\caption{$\varepsilon = 2.9,\, \Omega_s =(-20\pi,20\pi)^2,\, \Omega_c= (-1,1),\, \beta_s = 1.9,\, B = 3\pi,\, A = 9.$ Equilibria as functions of the nonlinear gain $\gamma$. The stable parts of the branches are indicated with thick lines. The dots indicate bifurcation points. Only the first three primary bifurcation points are shown. The right part of the Figure shows a zoomed version of the diagram in which the snaking is more apparent. Three images are shown at the bottom of the Figure, corresponding to the black stars in the two diagrams above. The colour surrounding the images corresponds to the one of its branch. }
		\label{fig:bdiag3a}
	\end{figure}

	We show  the results in Figure~\ref{fig:bdiag3a} where we have a snaking branch (green) of solutions arising from a 4D bifurcation point on the branch of stripes (red). We observe that the localization of the pattern is parallel to the stripes. Interestingly, there is another branch (data not shown) of patterns where the localization occurs perpendicular to the stripes and the branch  does not snake as explained in \cite{avitabile-lloyd-etal:10}.
	
	\begin{figure}[tbph!]
		\centering
		\includegraphics[width=0.8\textwidth]{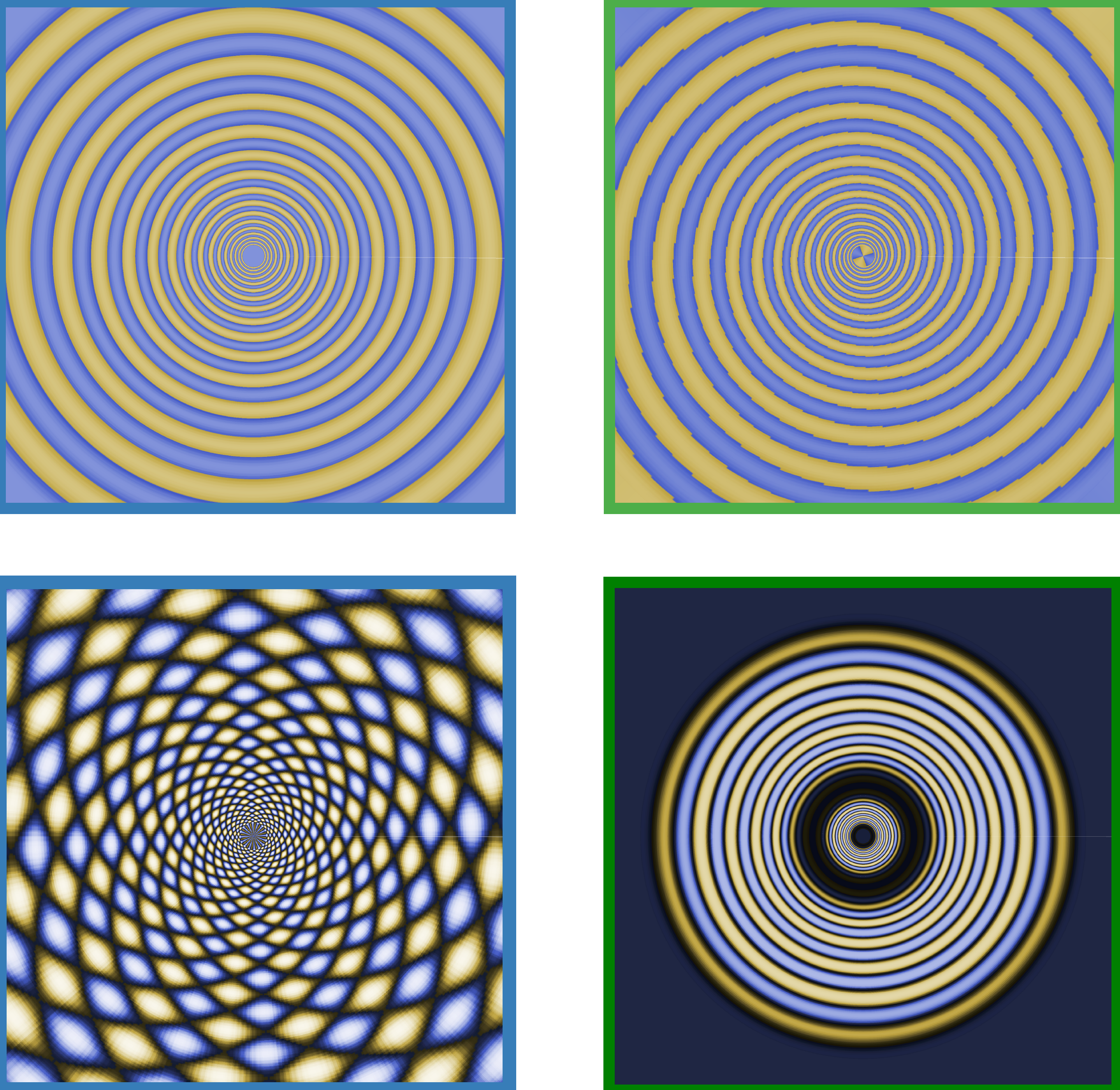}
		\caption{Stable equilibria in retinal coordinates. The top two images correspond to the first two in Figure~\ref{fig:bdiag1}. The bottom left image corresponds to the spot pattern (second one) in Figure~\ref{fig:bdiag3}. The bottom right image corresponds to the rightmost image in Figure~\ref{fig:bdiag3a}. Note that the framing colors match.}
		\label{fig:stableRetinal}
	\end{figure}
	\newpage
	
	\section{Conclusion}\label{sect:conclusion}
	
	We have studied some aspects of the bifurcations of the solutions to the neural-field equations describing a model of the primary visual area V1. We had proposed a variant of this model in \cite{song-faugeras-etal:19} and validated its predictions with some psychovisual data. Here we have focused on the stationary solutions to our equations and their bifurcations which we loosely interpreted as being possible metaphors of visual hallucinations. Visual hallucinations in color are not very well documented in the literature and it is therefore very difficult to compare our model predictions with experimental observations.
	
	The closest work to ours is described in \cite{bressloff-cowan-etal:01} and several of the follow-up papers. In these authors' work there was no attempt to model color perception, the focus was on achromatic (i.e. black and white) vision and the interplay between space and edges (borders of objects in the scene) orientations. Like us they studied the bifurcations of the solutions to the neural field equations describing their model and proposed to interpret them as accounting for visual hallucinations. Just like ours, theirs was a rather loose metaphor.
	
	Going into the mathematics behind the model, we have established here and in \cite{veltz-faugeras:10,song-faugeras-etal:19} the well-posedness of this class of equations whose solutions are defined on a four-dimensional compact spatio-chromatic space representing part of the cortical organisation of the visual cortex. The group of symmetries that acts on this space arises naturally from its spatial extension (assumed to be a square) and the psychophysics of Hering's color perception theory that emphasizes the symmetry with respect to the origin in chromaticity space (corresponding to what is known as opponent colors). Our model equations are equivariant w.r.t to this spatio-chromatic group and, by identifying some of its axial subgroups we have been able to prove (in part numerically) using the equivariant branching lemma that the simplest stationary solution (which turns out to be 0) bifurcates at pitchforks from which arise branches of stationary solutions enjoying the symmetries of the corresponding axial subgroup.
	
	Going into the numerics, the Julia package developed by the third author has allowed us to explore a much larger range of values of the bifurcation parameter than for example the authors of \cite{bressloff-cowan-etal:01} whose approach, based on the use of normal forms, is not well-suited to the analysis of subcritical bifurcations. It has also allowed us to discover cases of snaking and hence to observe stable localised solutions. To our knowledge this is the first time this has been reported in a model of the primary visual area. It raises the fascinating  question of whether such spatio-chromatic patterns can be observed in animal or human perception.
	
	Going further into the numerics, this is to our knowledge the first time that a bifurcation analysis of an infinite dimensional system of integro-differential equations is entirely carried out on GPUs. By considerably reducing the response time of our computer simulations it has allowed us to explore in much greater depth than what had been possible before the complicated structure of the solutions to our model equations with the result that we will in the future be able to better confront our predictions with perceptual experiments.

	\appendix
	\section{Color space}\label{section:color space}
	It is not important for this work to precisely define the color space we use because the phenomena described in this paper are robust to changes of this definition as long as the color space can be thought of as the product of a luminance axis with a 2D opponent chromaticity space which is assumed to be a disc of radius one centered at the origin. There are many possible choices for the details of such space. For convenience we have adopted the HSL space for Hue, Saturation and Luminance \cite{joblove-greenberg:78}. The (Hue, Saturation) coordinates are precisely the polar coordinates in the 2D chromaticity space, $2 \pi \varphi$ is the polar angle, varying between $0$ and $2\pi$ since $\varphi$ varies between 0 and 1, and $\rho$ is the distance to the origin, varying between 0 and 1. 
	In Section~\ref{sec:hallucinations}  for minimizing computer memory we restrict ourselves to a diameter of the unit disc, i.e. $\varphi=\varphi_0=\frac 1 8$. The conversion from the HSL space to the usual RGB space is nonlinear and can be found in textbooks such as \cite{wyszecki-stiles:67,joblove-greenberg:78,koenderink:10}.

	\section*{Acknowledgments}
	The authors thank Pascal Chossat for useful discussions.
	They are also grateful to the OPAL infrastructure from Université Côte d'Azur for providing resources and support.
	\\
	The third author would like to thank Hannes Uecker for his help.
	%
	%
	%
	%
	
\def\bysame{\leavevmode ---------\thinspace}
\makeatletter\if@francais\providecommand{\og}{<<~}\providecommand{\fg}{~>>}
\else\gdef\og{``}\gdef\fg{''}\fi\makeatother
\def\cdrandname{\&}
\providecommand\cdrnumero{no.~}
\providecommand{\cdredsname}{eds.}
\providecommand{\cdredname}{ed.}
\providecommand{\cdrchapname}{chap.}
\providecommand{\cdrmastersthesisname}{Memoir}
\providecommand{\cdrphdthesisname}{PhD Thesis}

\end{document}